\documentclass[prd,floatfix,onecolumn,amsmath,amssymb,floatfix]{revtex4}
\usepackage{graphicx,color,dcolumn,booktabs,bm}
\usepackage{subfigure}
\usepackage{longtable,lscape}
\usepackage{amssymb}
\usepackage{indentfirst}
\usepackage{epsfig}
\usepackage{feynmf}   
\usepackage{slashed}  
\usepackage{cases}
\definecolor{maroon}{RGB}{139,25,150}
\usepackage{multirow}
\usepackage{float}
\usepackage{graphicx,color,dcolumn,booktabs,bm}
\usepackage[colorlinks, citecolor=blue,anchorcolor=red,menucolor=red, linkcolor=red,filecolor=red,runcolor=red,urlcolor=blue,frenchlinks=red, urlcolor=blue]{hyperref}
\usepackage{orcidlink}

\begin{document}
\preprint{}
\preprint{}
\title{\color{maroon}{Semileptonic and nonleptonic weak decays of bottom baryons $\Omega^{(*)}_{b}$}}

\author{L.~Khajouei$^{a}$\orcidlink{0000-0003-0181-1879}}
\author{K.~Azizi$^{a,b}$\orcidlink{0000-0003-3741-2167}} 
\email{kazem.azizi@ut.ac.ir} \thanks{Corresponding author} 

\affiliation{
	$^{a}$Department of Physics, University of Tehran, North Karegar Avenue, Tehran 14395-547, Iran\\
	$^{b}$Department of Physics, Dogus University, Dudullu-\"{U}mraniye, 34775 Istanbul, T\"{u}rkiye}

\date{\today}

\begin{abstract}
We present an investigation into the semileptonic and nonleptonic weak decays of bottom baryons $\Omega^{*}_{b}$ and $\Omega_{b}$ within the framework of three-point QCD sum rules. In the semileptonic sector, the $\Omega^{*}_b\rightarrow\Omega_c\ell\bar{\nu}_{\ell}$ and $\Omega_b\rightarrow\Omega^*_c\ell\bar{\nu}_{\ell}$ transitions are specifically considered.  Utilizing the operator product expansion up to dimension six, the responsible form factors of these decays are obtained. The acquired form factors enable us to determine the decay widths in three leptonic channels. Branching ratios related to the $\Omega_{b}$ baryon semileptonic decays are also presented. These invariant form factors are subsequently employed as inputs to determine the nonleptonic weak decay widths in various modes with emitting a pseudoscalar or vector meson. An extensive investigation into all possible decay channels of bottom baryons provides valuable information for future experiments to examine the SM predictions, explores the new physics effects in heavy baryonic decays, and advances  the understanding of the internal structure of heavy baryons.     
\end{abstract}
\maketitle
\section{Introduction}\label{intro}
The existence of the heavy baryons with at least one b heavy quark in their quark contents has been accurately predicted by the quark model. In light of great advancements in various experiments, some of these b-heavy baryons have been identified regarding their quantum numbers. A detailed analysis of the bottom baryons' properties, theoretically and experimentally, presents a stimulating environment to probe the hadronic structures and the dynamics of strong interactions. Among the properties of b-heavy baryons, their decay modes draw great attention. The study of weak decays of the bottom baryons provides an ideal research field to examine the standard model (SM) of particle physics  and assess the new physics effects beyond the SM (BSM). An extensive investigation into the weak transitions as well as radiative and strong decays is crucial to gain a greater quantitative understanding of the main features of the quantum chromodynamics (QCD), in perturbative and nonperturbative sectors. Regarding the color confinement, determination of the CKM matrix elements is not directly possible, and semileptonic weak decays have a pivotal role in presenting the precise values of these matrix elements. In addition, the semileptonic weak processes offer a comprehensive insight into vector-axial vector structure. The nonleptonic weak decays are particularly prominent in a detailed study of CP violation.

In this study, we discuss the semileptonic and nonleptonic weak decays of singly b-heavy baryons,  $\Omega^*_b$ and $\Omega_b$, of different spin-$\frac{3}{2}$ and spin-$\frac{1}{2}$ states, respectively, in various decay channels. In particular, we consider the semileptonic and nonleptonic weak decays of $\Omega^{*}_b\to\Omega_c$ and $\Omega_b\to\Omega^*_c$ transitions. In 2008, the $\Omega_b$ baryon was identified by D$0$ Collaboration \cite{D0:2005cnn} in the decay channel $\Omega^-_b\to J/\Psi\Omega^-$ where $J/\Psi\to\mu^+\mu^-$, $\Omega^-\to\Lambda$K$^-$ and $\Lambda\to p\pi^-$ at an integrated luminosity of 1.3 fb$^{-1}$ and $\sqrt{s}$=1.96 TeV in $p\bar p$ collision \cite{D0:2008sbw}. In 2009, the observation of $\Omega_b$ was confirmed by CDF \cite{CDF:2004jtw} Collaboration \cite{CDF:2009sbo}. This heavy baryon was also reconstructed from the $\Omega^-_b\to J/\Psi\Omega^-$ decay, using 1.0 fb$^{-1}$ of data recorded in 2011 with the LHCb detector \cite{LHCb:2013wmn}. Moreover, the $\Omega_b$ baryon was established in another decay mode, $\Omega^-_b\to\Omega^0_c \pi^-$,  $\Omega^0_c\to p$K$^-$K$^-\pi^+$ at $\sqrt{s}$=7 and 8 TeV by the LHCb experiment \cite{LHCb:2016coe}.  $\Omega_b(6316)$,  $\Omega_b(6330)$,  $\Omega_b(6340)$, and  $\Omega_b(6350)$, as excited states of this heavy baryon, were identified at center of mass energies of 7, 8, and 13 TeV, related to a total integrated luminosity of 9 fb$^{-1}$ collected by the LHCb detector \cite{LHCb:2020tqd}. BABAR Collaboration found the $\Omega^*_c$ state through its radiative transition $\Omega^0_c \gamma$, with $\Omega^0_c$ reconstructed in the four decay channels $\Omega^-\pi^+$, $\Omega^-\pi^+\pi^0$, $\Omega^-\pi^+\pi^-\pi^+$, and $\Xi^-K^-\pi^+\pi^+$ \cite{BaBar:2006pve}. $\Omega^0_c$ was identified in numerous decay modes by a variety of Collaborations \cite{E687:1992tqn, BaBar:2007jdg}.\\An extensive range of theoretical approaches have been employed to investigate the semileptonic and nonleptonic weak decays of heavy baryons, including lattice QCD \cite{Bahtiyar:2021voz, Meinel:2016dqj, Detmold:2015aaa,Meinel:2021rbm}, light-front quark model \cite{Zhao:2022vfr, Geng:2020gjh, Ke:2019smy, Lu:2023rmq, Li:2021kfb, Ke:2024aux}, light cone QCD sum rules \cite{Hu:2019bqj, Aliev:2021wat, Aliev:2023tpk}, relativistic quark model \cite{Faustov:2018ahb}, constituent quark model \cite{Wang:2022zja, Wang:2024ozz, Dahiya:2024ekj}, nonrelativistic quark model \cite{Sharma:2009zzj, Li:2025mxz}, covariant confined quark model \cite{Gutsche:2015mxa}. We conduct a study of weak decays of the singly b-heavy baryons,  $\Omega^*_b$ and $\Omega_b$, within a predictive and effective nonperturbative method in hadronic physics, QCD sum rules, and compute their semileptonic and nonleptonic decay widths across different decay channels. QCD sum rule method, as a well-established approach having  compatible results with the experiments, is widely employed to determine various hadronic parameters \cite{Agaev:2017jyt, Aliev:2010uy,Khodjamirian:2011jp, Agaev:2016dev, Azizi:2016dhy}. In this framework, the hadron-to-hadron transition matrix element is parametrized in respect of the fundamental quantities called form factors. A detailed investigation into the form factors of different processes provides us with a profound insight into the internal structure of hadrons. We compute the responsible form factors and decay widths of the semileptonic $\Omega^{*}_b\rightarrow\Omega_c\ell\bar{\nu}_{\ell}$ and $\Omega_b\rightarrow\Omega^*_c\ell\bar{\nu}_{\ell}$ weak transitions in three lepton channels. These form factors are subsequently utilized as inputs into the computation of nonleptonic weak decay widths in different modes.

In the following, we derive the sum rules associated with the form factors of the semileptonic  $\Omega^{*}_b\rightarrow\Omega_c\ell\bar{\nu}_{\ell}$ transition, employing QCD sum rule method, in Sec. \ref{sum rules}. Sec. \ref{numerical} is devoted to the numerical analysis of the ultimate sum rules for the form factors through specifying the practical areas of the auxiliary parameters. In Sec. \ref{decayA}, using the fit functions of the form factors, the decay widths of the $\Omega^{*}_b\rightarrow\Omega_c$ semileptonic (nonleptonic) transitions are computed. We present the decay widths and branching ratios of the $\Omega_b\rightarrow\Omega^{*}_c$ semileptonic (nonleptonic) transitions in Sec. \ref{decayB}. Concluding remarks are given in Sec \ref{conclusion}. Appendix \ref{AppA} is prepared to give complementary calculations of the semileptonic $\Omega_b\rightarrow\Omega^*_c\ell\bar{\nu}_{\ell}$ transition, including the explicit expressions of the physical and QCD side. Appendix \ref{AppB} provides the representations of the sum rules associated with the form factors of the semileptonic $\Omega^{*}_b\rightarrow\Omega_c\ell\bar{\nu}_{\ell}$ transition. Further details of the computations are given in appendix \ref{AppC}.        
\section{QCD SUM RULES METHOD}\label{sum rules} 
Determination of various hadronic parameters, in order to probe the hadronic structures, is of interest for different theoretical approaches in particle physics. QCD sum rules, as a nonperturbative method, is employed to present a  description of the internal structure of hadrons \cite{Shifman:2010zzb, Shifman:1978by, Shifman:1978bx, Gross:2022hyw, Khodjamirian:2020btr, Colangelo:2000dp}. The main characteristic of this framework is being based on the QCD Lagrangian. In QCD sum rules, a correlation function, as a significant feature of this method, is computed in two discrete sides. One side is related to the short-distance quark-gluon interactions and is represented, utilizing operator product expansion (OPE), in terms of the quark and gluon degrees of freedom.  On the other side, long-distance interactions are dominant and the correlation function is described in terms of hadronic degrees of freedom. The latter is named physical or phenomenological side and the former is known as QCD or theoretical side. The physical quantities are derived from matching the coefficients of corresponding Lorentz structures from both sides.\\In the present study, we concentrate on semileptonic and nonleptonic weak decays of singly b-heavy baryons, $\Omega^*_b$ and $\Omega_b$. We acquire the responsible form factors of the $\Omega^{*}_b\rightarrow\Omega_c\ell\bar{\nu}_{\ell}$ and $\Omega_b\rightarrow\Omega^*_c\ell\bar{\nu}_{\ell}$ semileptonic weak transitions and calculate the decay widths in all leptonic channels. The obtained results are subsequently applied to an investigation of nonleptonic weak decays of these heavy baryons.  In the following, we consider the $\Omega^{*}_b\rightarrow\Omega_c\ell\bar{\nu}_{\ell}$ transition and obtain its corresponding correlation function on both phenomenological and QCD sides. The explicit expressions of the calculations related to the other transition,  $\Omega_b\rightarrow\Omega^*_c\ell\bar{\nu}_{\ell}$, are provided in appendix \ref{AppA}.

The semileptonic $\Omega^{*}_b\rightarrow\Omega_c\ell\bar{\nu}_{\ell}$ weak decay is induced by the $b\rightarrow c\ell\bar{\nu}_{\ell}$ transition at the quark level. Indeed, an effective lepton-quark interaction, via the exchange of a massive W-boson, is regarded as the basis of this semileptonic weak transition and can be expressed by the effective Hamiltonian,
\begin{equation}
	\mathcal{H}_{eff}=\frac{G_{F}}{\sqrt{2}}V_{cb}\,\bar{c}\gamma_{\mu}(1-\gamma_{5})b\,\bar{\ell}\gamma^{\mu}(1-\gamma_{5})\nu_{\ell},
\end{equation}
where $G_{F}$ denotes the Fermi coupling constant and $V_{cb}$ represents the CKM matrix element, involved in $b\rightarrow c$ transition. The decay amplitude is evaluated by putting the effective Hamiltonian between the initial and final baryonic states,
 \begin{equation}
	M=\langle\Omega_{c}|\mathcal{H}_{eff}|\Omega_{b}^{*}\rangle,
\end{equation}
in which just the quark part remains in the matrix element,  $\langle\Omega_{c}|\bar{c}\gamma_{\mu}(1-\gamma_{5})b|\Omega^{*}_{b}\rangle$, and,  considering  the Lorentz invariance and parity considerations, it is parametrized as,
\begin{eqnarray}\label{formfactor}
	&&\langle\Omega_{c}(p^{\prime},s^{\prime})|J_{\mu}^{V}|\Omega_{b}^{*}(p,s)\rangle=
	\langle\Omega_{c}(p^{\prime},s^{\prime})|\bar{c}\gamma_{\mu}b|\Omega_{b}^{*}(p,s)\rangle\nonumber\\
	&&=\bar{u}_{\Omega_{c}}(p^{\prime},s^{\prime})\Big[g_{\mu\alpha}F_{1}(q^{2})+\gamma_{\mu}\frac{p^{\prime}_{\alpha}}{M_{\Omega_{c}}}F_{2}(q^{2})+\frac{p^{\prime}_{\alpha}p_{\mu}}{M_{\Omega_{c}}^{2}}F_{3}(q^{2})+\frac{p^{\prime}_{\alpha}q_{\mu}}{M_{\Omega_{c}}^{2}}F_{4}(q^{2})\Big]\gamma_{5}u_{\Omega_{b}^{*}}^{\alpha}(p,s),\nonumber\\
	&&\nonumber\\
	&&\langle\Omega_{c}(p^{\prime},s^{\prime})|J_{\mu}^{A}|\Omega_{b}^{*}(p,s)\rangle=
	\langle\Omega_{c}(p^{\prime},s^{\prime})|\bar{c}\gamma_{\mu}\gamma_{5}b|\Omega_{b}^{*}(p,s)\rangle\nonumber\\
	&&=\bar{u}_{\Omega_{c}}(p^{\prime},s^{\prime})\,\Big[g_{\mu\alpha}G_{1}(q^{2})+\gamma_{\mu}\frac{p^{\prime}_{\alpha}}{M_{\Omega_{c}}}\,G_{2}(q^{2})+\frac{p^{\prime}_{\alpha}p_{\mu}}{M_{\Omega_{c}}^{2}}G_{3}(q^{2})+\frac{p^{\prime}_{\alpha}q_{\mu}}{M_{\Omega_{c}}^{2}}G_{4}(q^{2})\Big]u_{\Omega_{b}^{*}}^{\alpha}(p,s),
\end{eqnarray}	
where $F_{1}(q^2), F_{2}(q^2), F_{3}(q^2)$, and $F_{4}(q^2)$ are the form factors related to the vector part of the weak transition current, $J_{\mu}^{V}$, and $G_{1}(q^2), G_{2}(q^2), G_{3}(q^2)$, and $G_{4}(q^2)$ correspond to the axial vector part of the weak current, $J_{\mu}^{A}$. $q=p-p'$ represents the momentum transferred to the leptonic part, i.e., lepton and corresponding antineutrino. $u_{\Omega_{b}^{*}}^{\alpha}(p, s)$ indicates the Rarita-Schwinger spinor of the initial baryonic state with momentum $p$ and spin $s$, while the Dirac spinor associated with the final baryon with momentum $p'$ and spin $s'$ is denoted as  $u_{\Omega_{c}}(p',s')$.

We proceed with the computation of the responsible form factors of the  $\Omega^{*}_b\rightarrow\Omega_c\ell\bar{\nu}_{\ell}$ transition via the three-point QCD sum rule method. In this approach, the correlation function represents a baryonic transition, consisting of the initial and final baryonic currents connected with a transition one. Hence, the subsequent correlation function is considered,
 \begin{equation}\label{correlation}
	\Pi_{\mu\nu}(p,p^{\prime},q^2)=i^{2}\int\mathrm{d}^{4}x e^{-ip.x}\int\mathrm{d}^{4}y e^{ip^{\prime}.y}\langle 0|{\cal T}\{{\cal J}^{\Omega_{c}}(y){\cal J}_{\mu}^{tr}(0)\bar{\cal J}_{\nu}^{\Omega_{b}^{*}}(x)\}|0\rangle,
\end{equation} 
where ${\cal T}$ denotes the time-ordering operator, the interpolating currents of the initial and final states are represented as ${\cal J}^{\Omega_{b}^{*}}$ and ${\cal J}^{\Omega_{c}}$, respectively, and ${\cal J}_{\mu}^{tr}$ is the transition current between the $\Omega_{b}^{*}$, a singly b-heavy baryon with spin $\frac{3}{2}$, and the $\Omega_{c}$, a singly c-heavy baryon with spin $\frac{1}{2}$, in $\frac{3}{2}\rightarrow\frac{1}{2}$ transition (See Table \ref{omega}). 
 \begin{table}[h!]
	\centering
	\begin{tabular}{ccccc}
		\hline
		Particle~~~&Quark constituents~~~&Charge~~~&Quark model~~~&Spin parity~\\
		\hline
		$\Omega_{b}^{*}$&$(ssb)$&-1&sextet&$\frac{3}{2}^{+}$\\
		$\Omega_{c}$&$(ssc)$&0&sextet&$\frac{1}{2}^{+}$\\
		$\Omega_{b}$&$(ssb)$&-1&sextet&$\frac{1}{2}^{+}$\\
		$\Omega_{c}^{*}$&$(ssc)$&0&sextet&$\frac{3}{2}^{+}$\\
		\hline
		\hline
	\end{tabular}
	\caption{ 
		Quantum numbers and quark constituents of $\Omega_{b}^{(*)}$ and $\Omega_{c}^{(*)}$ baryons.}
		\label{omega}
\end{table}
\subsection{Phenomenological representation}
With regard to the complicated content of the correlation function, including ground states, excited, and a continuum of many-particle hadronic states, a quantitative correlation function is achieved by embedding two complete sets of the hadronic modes, whose quantum numbers and properties are consistent with the interpolating currents of the initial and final hadronic states, into the expression of the correlation function \cite{Khodjamirian:2020btr, Colangelo:2000dp}, Eq. \eqref{correlation},
 \begin{eqnarray}
	1=\vert 0\rangle \langle 0\vert+\sum_{h}\int\frac{\mathrm{d}^{4}p_{h}}{(2\pi)^{4}}2\pi\delta(p_{h}^{2}-m_{h}^{2})\vert h(p_{h})\rangle\langle h(p_{h})\vert +\text{higher Fock states},
\end{eqnarray}
where $\vert h(p_h)\rangle$ represents the likely hadronic state with momentum $p_h$.\\The physical side of the correlation function is acquired, by entering the complete sets in the Eq. \eqref{correlation} and employing the Fourier transformation and integrating over x and y in four-dimension, as,
   \begin{eqnarray}\label{physical}
  	\Pi_{\mu\nu}^{Phys.}(p,p^{\prime},q^{2})=\frac{\langle 0\mid{\cal J}^{\Omega_{c}}(0)\mid\Omega_{c}(p^{\prime})\rangle\langle\Omega_{c}(p^{\prime})\mid{\cal J}_{\mu}^{tr}(0)\mid\Omega_{b}^{*}(p)\rangle\langle\Omega_{b}^{*}(p)\mid\bar{\cal J}_{\nu}^{\Omega_{b}^{*}}(0)\mid 0\rangle}{(p^{\prime2}-m_{\Omega_{c}}^{2})(p^{2}-m_{\Omega_{b}^{*}}^{2})}+\cdots ,
  \end{eqnarray} 
  where ellipsis indicates the contributions of the higher states and continuum. After introducing the residue of the initial baryonic state, $\lambda_{\Omega_b^*}$, and final  state, $\lambda_{\Omega_c}$, as,
   \begin{eqnarray}\label{residue}
  	\langle0|{\cal J}^{\Omega_{c}}(0)|\Omega_{c}(p^{\prime})\rangle&=&\lambda_{\Omega_{c}}u_{\Omega_{c}}(p^{\prime},s^{\prime}),\notag\\
  	\langle\Omega_{b}^{*}(p)|\bar{\cal J}_{\nu}^{\Omega_{b}^{*}}(0)|0\rangle&=&\lambda_{\Omega_{b}^{*}}^{\dagger}\bar{u}_{\nu\Omega_{b}^{*}}(p,s),
  \end{eqnarray}
  and substituting the Eqs. \eqref{formfactor}, and \eqref{residue} in Eq. \eqref{physical},  
we apply the summation relation over Dirac spinors with spin $\frac{1}{2}$,
 \begin{equation}
	\sum_{s^{\prime}}u_{\Omega_{c}}(p^{\prime},s^{\prime})\bar{u}_{\Omega_{c}}(p^{\prime},s^{\prime})=(\slashed{p}^{\prime}+m_{\Omega_{c}}),
\end{equation}    
  and Rarita-Schwinger spinors with spin $\frac{3}{2}$, 
   \begin{equation}
  	\sum_{s}u_{\alpha\,\Omega_{b}^{*}}(p,s)\bar{u}_{\nu\Omega_{b}^{*}}(p,s)=-(\slashed{p}+m_{\Omega_{b}^{*}})\Big(g_{\alpha\nu}-\frac{\gamma_{\alpha}\gamma_{\nu}}{3}-\frac{2}{3}\frac{p_{\alpha}p_{\nu}}{m_{\Omega_{b}^{*}}^{2}}+\frac{1}{3}\frac{p_{\alpha}\gamma_{\nu}-p_{\nu}\gamma_{\alpha}}{m_{\Omega_{b}^{*}}}\Big).
  \end{equation}
  to obtain the phenomenological representation of the correlation function. Here, it is essential to mention some remarks. Since the Lorentz structures of the correlation function are not completely independent, it is required to arrange the matrices in a particular order. In this study, we utilize an ordering of the Lorentz structures as $\gamma_{\mu}\slashed{p}\slashed {p}^{\prime}\gamma_{\nu}\gamma_{5}$. Correlation function includes the contributions from both spin-$\frac{1}{2}$ and spin-$\frac{3}{2}$ states. It originates from the nature of the interpolating current of spin-$\frac{3}{2}$ state which can additionally produce spine-$\frac{1}{2}$ states from vacuum. These contributions, employing the $\gamma^{\nu}{\cal J}_{\nu}=0$ constraint, are expressed as follows,
   \begin{equation}
   	 \langle0|{\cal J}_{\nu}^{\Omega_{b}^{*}}|(p,s=\frac{1}{2})\rangle=(\alpha\gamma_{\nu}-\frac{4\alpha}{m} p_{\nu})u(p,s=\frac{1}{2}).
  \end{equation}  
  It reveals that the structures $\gamma_{\nu}$ and $p_{\nu}$ incorporate the contributions of spin-$\frac{1}{2}$ states. In order to eliminate these contributions, we remove the structures in proportion to $\gamma_{\nu}$ and $p_{\nu}$.\\As mentioned previously, the correlation function incorporates the contributions of higher states and continuum. With the purpose of suppressing these terms, three-point QCD sum rules method employs the Borel transformation and continuum subtraction which bring the auxiliary parameters to the method; the Borel mass parameters, ($M^2, M^{\prime 2}$), and the continuum thresholds, ($s_0, s^{\prime}_0$). The other parameter is a mathematical mixing one, $\beta$, arising from the spin-$\frac{1}{2}$ interpolating current. The working areas of these parameters, ensuring the stability of physical quantities, are fixed and discussed in  Sec. \ref{numerical}. Utilizing the double Borel transformation,
   \begin{equation}\label{Borel transformation} 
  	\mathbf{\widehat{B}}\frac{1}{(p^2-m^{2})^{m}}\frac{1}{(p^{\prime2}-m^{\prime2})^{n}}\longrightarrow(-1)^{m+n}\frac{1}{\Gamma[m]\Gamma[n]}\frac{1}{(M^{2})^{m-1}}\frac{1}{(M^{\prime2})^{n-1}}e^{-m^{2}/M^{2}}e^{-m^{\prime2}/M^{\prime2}},
  \end{equation}
  where $M^2$ and $M^{'2}$ are Borel parameters and $m$ and $m'$ are the masses of the initial and final states, respectively, we obtain the physical or phenomenological representation of the correlation function as,
  \begin{eqnarray}\label{physicalside}
  	&&\mathbf{\widehat{B}}\Pi_{\mu\nu}^{Phys.}(p,p^{\prime},q^2)=\lambda_{\Omega_{b}^{*}}\lambda_{\Omega_{c}}e^{-\frac{m_{\Omega_{b}^{*}}^{2}}{M^{2}}}e^{-\frac{m_{\Omega_{c}}^{2}}{M\prime^{2}}}\Bigg[F_{1}\Big(-m_{\Omega_{b}^{*}}m_{\Omega_{c}}g_{\mu\nu}\gamma_{5}+m_{\Omega_{c}}g_{\mu\nu}\slashed{p}\gamma_{5}-m_{\Omega_{b}^{*}}g_{\mu\nu}\slashed{p}^{\prime}\gamma_{5}-g_{\mu\nu}\slashed{p}\slashed{p}^{\prime}\gamma_{5}\Big)+\nonumber\\&&F_{2}\Big(\frac{2}{m_{\Omega_{c}}}p^{\prime}_{\mu}p^{\prime}_{\nu}\slashed{p}\gamma_{5}+p^{\prime}_{\nu}\gamma_{\mu}\slashed{p}\gamma_{5}+\frac{m_{\Omega_{b}^{*}}}{m_{\Omega_{c}}}p^{\prime}_{\nu}\gamma_{\mu}\slashed{p}^{\prime}\gamma_{5}+\frac{1}{m_{\Omega_{c}}}p^{\prime}_{\nu}\gamma_{\mu}\slashed{p}\slashed{p}^{\prime}\gamma_{5}\Big)+F_{3}\Big(-\frac{m_{\Omega_{b}^{*}}}{m_{\Omega_{c}}}p_{\mu}p^{\prime}_{\nu}\gamma_{5}+\frac{1}{m_{\Omega_{c}}}p_{\mu}p^{\prime}_{\nu}\slashed{p}\gamma_{5}-\nonumber\\&&\frac{m_{\Omega_{b}^{*}}}{m_{\Omega_{c}}^{2}}p_{\mu}p^{\prime}_{\nu}\slashed{p}^{\prime}\gamma_{5}-\frac{1}{m_{\Omega_{c}}^{2}}p_{\mu}p^{\prime}_{\nu}\slashed{p}\slashed{p}^{\prime}\gamma_{5}\Big)+F_{4}\Big(\frac{m_{\Omega_{b}^{*}}}{m_{\Omega_{c}}^{2}}p^{\prime}_{\mu}p^{\prime}_{\nu}\slashed{p}^{\prime}\gamma_{5}-\frac{m_{\Omega_{b}^{*}}}{m_{\Omega_{c}}^{2}}p_{\mu}p^{\prime}_{\nu}\slashed{p}^{\prime}\gamma_{5}-\frac{1}{m_{\Omega_{c}}^{2}}p_{\mu}p^{\prime}_{\nu}\slashed{p}\slashed{p}^{\prime}\gamma_{5}+\frac{1}{m_{\Omega_{c}}^{2}}p^{\prime}_{\mu}p^{\prime}_{\nu}\slashed{p}\slashed{p}^{\prime}\gamma_{5}\Big)\nonumber\\&&-G_{1}\Big(-m_{\Omega_{b}^{*}}m_{\Omega_{c}}g_{\mu\nu}-m_{\Omega_{c}}g_{\mu\nu}\slashed{p}-m_{\Omega_{b}^{*}}g_{\mu\nu}\slashed{p}^{\prime}+g_{\mu\nu}\slashed{p}\slashed{p}^{\prime}\Big)-G_{2}\Big(-\frac{2}{m_{\Omega_{c}}}p^{\prime}_{\mu}p^{\prime}_{\nu}\slashed{p}-p^{\prime}_{\nu}\gamma_{\mu}\slashed{p}+\frac{m_{\Omega_{b}^{*}}}{m_{\Omega_{c}}}p^{\prime}_{\nu}\gamma_{\mu}\slashed{p}^{\prime}\nonumber\\&&-\frac{1}{m_{\Omega_{c}}}p^{\prime}_{\nu}\,\gamma_{\mu}\slashed{p}\slashed{p}^{\prime}\Big)-G_{3}\Big(-\frac{m_{\Omega_{b}^{*}}}{m_{\Omega_{c}}}p_{\mu}p^{\prime}_{\nu}-\frac{1}{m_{\Omega_{c}}}p_{\mu}p^{\prime}_{\nu}\slashed{p}-\frac{m_{\Omega_{b}^{*}}}{m_{\Omega_{c}}^{2}}p_{\mu}p^{\prime}_{\nu}\slashed{p}^{\prime}+\frac{1}{m_{\Omega_{c}}^{2}}p_{\mu}p^{\prime}_{\nu}\slashed{p}\slashed{p}^{\prime}\Big)-G_{4}\Big(\frac{m_{\Omega_{b}^{*}}}{m_{\Omega_{c}}^{2}}p^{\prime}_{\mu}p^{\prime}_{\nu}\slashed{p}^{\prime}\nonumber\\&&-\frac{m_{\Omega_{b}^{*}}}{m_{\Omega_{c}}^{2}}p_{\mu}p^{\prime}_{\nu}\slashed{p}^{\prime}+\frac{1}{m_{\Omega_{c}}^{2}}p_{\mu}p^{\prime}_{\nu}\slashed{p}\slashed{p}^{\prime}-\frac{1}{m_{\Omega_{c}}^{2}}p^{\prime}_{\mu}p^{\prime}_{\nu}\slashed{p}\slashed{p}^{\prime}\Big)\Bigg]+\cdots,
  \end{eqnarray}
  The form factors $F_i(q^2)$ and $G_i(q^2), (i=1,2,3,4)$ are Lorentz invariant and q$^2$- dependent. Here, for simplicity, the q$^2$ dependency is excluded.
  \subsection{QCD representation}
  To satisfy the requirement of the QCD sum rules approach, correlation function is also determined in the deep Euclidean region, $Q^{2}=-q^{2}\gg\Lambda_{QCD}^{2}$, in terms of quark and gluon degrees of freedom. In this region, regarding both perturbative and nonperturbative contributions, the OPE is utilized and correlation function is expressed, in respect of the  vacuum condensates, as,
   \begin{equation}
  	\Pi(q^{2})=\sum_{d}C_{d}(q^{2})\langle0\vert O_{d}\vert0\rangle,
  \end{equation}
  where $C_d(q^2)$ are Wilson coefficients and $O_d$ are operators with different dimensions, $d$, and the vacuum expectation values of the mentioned operators represent a series of vacuum condensates. Hence, to compute the QCD or theoretical side of the correlation function, in  three-point QCD sum rules method, the interpolating currents of the initial and final baryonic modes are employed. These currents, corresponding to the quark fields, are represented as,
   \begin{equation}\label{currentomegab}
  	{\cal J}_{\nu}^{\Omega_{b}^{*}}(x)=\sqrt\frac{1}{3}\epsilon^{abc}\Bigg\{ \Big(s^{aT}(x)C\gamma_{\nu}s^{b}(x)\Big)b^{c}(x)+\Big(s^{aT}(x)C\gamma_{\nu}b^{b}(x)\Big)s^{c}(x)+\Big(b^{aT}(x)C\gamma_{\nu}s^{b}(x)\Big)s^{c}(x)\Bigg\},
  \end{equation} 
  for the $\Omega_b^{*}$, the initial singly b-heavy baryon with spin $\frac{3}{2}$ and belonging to the $6_F$ representation of the quark model, and
   \begin{eqnarray}\label{currentomegac}
  	{\cal J}^{\Omega_{c}}(y)=-\frac{1}{\sqrt{2}}\epsilon^{abc}\Bigg\{ \Big(s^{aT}(y)Cc^{b}(y)\Big)\gamma_{5}s^{c}(y)+\beta\Big(s^{aT}(y)C\gamma_{5}c^{b}(y)\Big)s^{c}(y)\notag\\
  	-\Big(c^{aT}(y)Cs^{b}(y)\Big)\gamma_{5}s^{c}(y)-\beta\Big(c^{aT}(y)C\gamma_{5}s^{b}(y)\Big)s^{c}(y)\Bigg\},
  \end{eqnarray}
  for the final singly c-heavy baryon of spin $\frac{1}{2}$ and $6_F$ representation of the quark model, $\Omega_c$, where $a$, $b$, and $c$ are color indices, $C$ is the charge conjugation operator, and $b(x)$, $c(x)$, and $s(x)$ indicate the quark fields; details have been discussed in appendix A of Ref. \cite{Khajouei:2024frw}. Substituting these interpolating currents, given in Eqs. \eqref{currentomegab} and \eqref{currentomegac}, into the expression of the correlation function, Eq. \eqref{correlation}, and applying Wick's theorem, by contracting the related quark fields, the QCD side of the correlation function is derived as follows, 
  \begin{eqnarray}\label{qcdside}
  	&&\Pi_{\mu\nu}^{QCD}(p,p^{\prime },q^{2})=i^{2}\int\mathrm{d}^{4}x e^{-ipx}\int\mathrm{d}^{4}y e^{ip^{\prime}y}\frac{1}{\sqrt{2}}\frac{1}{\sqrt{3}}\epsilon^{abc}\epsilon^{a^{\prime} b^{\prime} c^{\prime}}\Bigg\{\gamma_5 S^{cb'}_{s}(y-x) \gamma_{\nu} \tilde{S}^{aa'}_{s}(y-x) S^{bi}_{c}(y) \gamma_{\mu}(1-\gamma_5)\nonumber\\&& S^{ic'}_{b}(-x)-\gamma_5 S^{ca'}_{s}(y-x) \gamma_{\nu} \tilde{S}^{ab'}_{s}(y-x) S^{bi}_{c}(y) \gamma_{\mu}(1-\gamma_5)S^{ic'}_{b}(-x)-\gamma_5 S^{cc'}_{s}(y-x) Tr[\gamma_{\nu} \tilde{S}^{aa'}_{s}(y-x) S^{bi}_{c}(y)\nonumber\\&&\gamma_{\mu}(1-\gamma_5) S^{ib'}_{b}(-x)]-\gamma_5 S^{ca'}_{s}(y-x)\gamma_{\nu}\tilde{S}^{ib'}_{b}(-x)(1-\gamma_5)\gamma_{\mu}\tilde{S}^{bi}_{c}(y)S^{ac'}_{s}(y-x)-\gamma_5 S^{cc'}_{s}(y-x)Tr[\gamma_{\nu}\tilde{S}^{ia'}_{b}(-x)\nonumber\\&&(1-\gamma_5)\gamma_{\mu}\tilde{S}^{bi}_{c}(y)S^{ab'}_{s}(y-x)]+\gamma_5 S^{cb'}_{s}(y-x)\gamma_{\nu}\tilde{S}^{ia'}_{b}(-x)(1-\gamma_5)\gamma_{\mu}\tilde{S}^{bi}_{c}(y)S^{ac'}_{s}(y-x)+\beta S^{cb'}_{s}(y-x)\gamma_{\nu}\nonumber\\&&\tilde{S}^{aa'}_{s}(y-x)\gamma_5 S^{bi}_{c}(y)\gamma_{\mu}(1-\gamma_5)S^{ic'}_{b}(-x)-\beta S^{ca'}_{s}(y-x)\gamma_{\nu}\tilde{S}^{ab'}_{s}(y-x)\gamma_5 S^{bi}_{c}(y)\gamma_{\mu}(1-\gamma_5)S^{ic'}_{b}(-x)-\beta\nonumber\\&& S^{cc'}_{s}(y-x)Tr[\gamma_{\nu}\tilde{S}^{aa'}_{s}(y-x)\gamma_5 S^{bi}_{c}(y)\gamma_{\mu}(1-\gamma_5)S^{ib'}_{b}(-x)]-\beta S^{ca'}_{s}(y-x)\gamma_{\nu}\tilde{S}^{ib'}_{b}(-x)(1-\gamma_5)\gamma_{\mu}\tilde{S}^{bi}_{c}(y)\gamma_5\nonumber\\&& S^{ac'}_{s}(y-x)-\beta S^{cc'}_{s}(y-x)Tr[\gamma_{\nu}\tilde{S}^{ia'}_{b}(-x)(1-\gamma_5)\gamma_{\mu}\tilde{S}^{bi}_{c}(y)\gamma_5 S^{ab'}_{s}(y-x)]+\beta S^{cb'}_{s}(y-x)\gamma_{\nu}\tilde{S}^{ia'}_{b}(-x)(1-\gamma_5)\nonumber\\&&\gamma_{\mu}\tilde{S}^{bi}_{c}(y)\gamma_5 S^{ac'}_{s}(y-x)-\gamma_5 S^{cb'}_{s}(y-x)\gamma_{\nu}\tilde{S}^{ba'}_{s}(y-x)S^{ai}_{c}(y)\gamma_{\mu}(1-\gamma_5)S^{ic'}_{b}(-x)+\gamma_5 S^{ca'}_{s}(y-x)\gamma_{\nu}\tilde{S}^{bb'}_{s}(y-x)\nonumber\\&&S^{ai}_{c}(y)\gamma_{\mu}(1-\gamma_5)S^{ic'}_{b}(-x)+\gamma_5 S^{cc'}_{s}(y-x)Tr[\gamma_{\nu}\tilde{S}^{ba'}_{s}(y-x)S^{ai}_{c}(y)\gamma_{\mu}(1-\gamma_5)S^{ib'}_{b}(-x)]+\gamma_5 S^{ca'}_{s}(y-x)\gamma_{\nu}\nonumber\\&&\tilde{S}^{ib'}_{b}(-x)(1-\gamma_5)\gamma_{\mu} \tilde{S}^{ai}_{c}(y)S^{bc'}_{s}(y-x)+\gamma_5 S^{cc'}_{s}(y-x)Tr[\gamma_{\nu}\tilde{S}^{ia'}_{b}(-x)(1-\gamma_5)\gamma_{\mu}\tilde{S}^{ai}_{c}(y)S^{bb'}_{s}(y-x)]-\gamma_5\nonumber\\&& S^{cb'}_{s}(y-x)\gamma_{\nu}\tilde{S}^{ia'}_{b}(-x)(1-\gamma_5)\gamma_{\mu}\tilde{S}^{ai}_{c}(y)S^{bc'}_{s}(y-x)-\beta S^{cb'}_{s}(y-x)\gamma_{\nu}\tilde{S}^{ba'}_{s}(y-x)\gamma_5 S^{ai}_{c}(y)\gamma_{\mu}(1-\gamma_5)\nonumber\\&&S^{ic'}_{b}(-x)+\beta S^{ca'}_{s}(y-x)\gamma_{\nu}\tilde{S}^{bb'}_{s}(y-x)\gamma_5 S^{ai}_{c}(y)\gamma_{\mu}(1-\gamma_5)S^{ic'}_{b}(-x)+\beta S^{cc'}_{s}(y-x)Tr[\gamma_{\nu}\tilde{S}^{ba'}_{s}(y-x)\gamma_5\nonumber\\&& S^{ai}_{c}(y)\gamma_{\mu}(1-\gamma_5)S^{ib'}_{b}(-x)]+\beta S^{ca'}_{s}(y-x)\gamma_{\nu}\tilde{S}^{ib'}_{b}(-x)(1-\gamma_5)\gamma_{\mu}\tilde{S}^{ai}_{c}(y)\gamma_5 S^{bc'}_{s}(y-x)+\beta S^{cc'}_{s}(y-x)\nonumber\\&&Tr[\gamma_{\nu}\tilde{S}^{ia'}_{b}(-x)(1-\gamma_5)\gamma_{\mu}\tilde{S}^{ai}_{c}(y)\gamma_5 S^{bb'}_{s}(y-x)]-\beta S^{cb'}_{s}(y-x)\gamma_{\nu}\tilde{S}^{ia'}_{b}(-x)(1-\gamma_5)\gamma_{\mu}\tilde{S}^{ai}_{c}(y)\gamma_5 S^{bc'}_{s}(y-x)\Bigg\},      
  \end{eqnarray}	
  where $\tilde{S_q}=CS_{q}^{T}C$. The explicit form of the light quark propagator is given as \cite{Agaev:2020zad},
   \begin{eqnarray}
  	&&S_{q}^{ab}(x)=
  	i\delta_{ab}\frac{\slashed{x}}{2\pi^{2}x^{4}}-\delta_{ab}\frac{m_{q}}{4\pi^{2}x^{2}}-\delta_{ab}\frac{\langle\bar{q}q\rangle}{12}
  	+i\delta_{ab}\frac{\slashed{x}m_{q}\langle\bar{q}q\rangle}{48}-\delta_{ab}\frac{x^{2}}{192}\langle\bar{q}g_{s}\sigma G q\rangle\nonumber\\&&+i\delta_{ab}\frac{x^{2}\slashed{x}m_{q}}{1152}<\bar{q}g_{s}\sigma G q>
  	-i\frac{g_{s}G_{ab}^{\mu\nu}}{32\pi^{2}x^{2}}[\slashed{x}\sigma_{\mu\nu}+\sigma_{\mu\nu}\slashed{x}]-i\delta_{ab}\frac{x^{2}\slashed{x}g_{s}^{2}\langle\bar{q}q\rangle^{2}}{7776}-\delta_{ab}\frac{x^{4}\langle\bar{q}q\rangle\langle g_{s}^{2}G^{2}\rangle}{27648}+\cdots ,
  \end{eqnarray} 
   where $\langle\bar{q}q\rangle$ represents the quark condensate, $\langle\bar{q}g_{s}\sigma G q\rangle$ corresponds to the quark-gluon condensate, and $m_q$ is the light quark mass. The heavy quark propagator, denoted as $S_{Q}^{ab}(x)$, is expressed through the next form \cite{Agaev:2020zad}, 
   \begin{eqnarray}
   	&&S_{Q}^{ab}(x)=
   	i\int\frac{\mathrm{d}^{4}k}{(2\pi)^{4}}e^{-ikx}\Bigg\{\frac{\delta_{ab}(\slashed{k}+m_{Q})}{k^{2}-m_{Q}^{2}}-\frac{g_{s}G_{ab}^{\mu\nu}}{4}\frac{\sigma_{\mu\nu}(\slashed{k}+m_{Q})+(\slashed{k}+m_{Q})\sigma_{\mu\nu}}{(k^{2}-m_{Q}^{2})^{2}}\nonumber\\
   	&&+\frac{g_{s}^{2}G^{2}}{12}\delta_{ab}m_{Q}\frac{k^{2}+m_{Q}\slashed{k}}{(k^{2}-m_{Q}^{2})^{4}}+\frac{g_{s}^{3}G^{3}}{48}\delta_{ab}\frac{(\slashed{k}+m_{Q})}{(k^{2}-m_{Q}^{2})^{6}}[\slashed{k}(k^{2}-3m_{Q}^{2})+2m_{Q}(2k^{2}-m_{Q}^{2})](\slashed{k}+m_{Q})+\cdots\Bigg\},
   \end{eqnarray}
  in terms of the heavy quark mass, $m_Q$, and the four-momentum of the heavy quark, $k$. In these terms, the following definitions are utilized,
   \begin{equation}
  	G_{ab}^{\mu\nu}\equiv G_{A}^{\mu\nu}t_{ab}^{A},~~G^{2}=G_{\mu\nu}^{A}G_{A}^{\mu\nu},~~t^{A}=\lambda^{A}/2,
  \end{equation}
  where $G^{\mu\nu}_{A}$ is the gluon field strength tensor, $\mu$ and $\nu$ indicate Lorentz indices, and $\lambda^A$ represents the Gell-Mann matrices with $A=1, 2, ..., 8$.
  The theoretical side of the correlation function incorporates both perturbative and nonperturbative contributions of different mass dimensions which arise from the operators of various dimensions in OPE; perturbative effects is attributed to the lowest-dimension operator, $d=0$, while nonperturbative effects are relevant to the vacuum condensates, including quark condensate of three dimension, $\langle \bar{q}q \rangle$, gluon and quark-gluon condensates,  $\langle G^{2}\rangle$ and $\langle\bar{q}g\sigma G q\rangle$, of four and five dimensions, respectively, and two-times quark condensate, related to the six dimension, $\langle \bar{q}q \rangle^{2}$. The computations of this study are performed up to six- dimension.
  
  To obtain the QCD side of the correlation function, it is required to compute the four-integrals of the form,
   \begin{equation}
  	\int \mathrm{d}^{4}k\int\mathrm{d}^{4}k^{\prime}\int \mathrm{d}^{4}x e^{i(k-p).x}\int \mathrm{d}^{4}y e^{i(-k^{\prime}+p^{\prime}).y}\frac{x_{\mu}y_{\nu}k_{\alpha}k^{\prime}_{\beta}}{(k^{2}-m_{b}^{2})^{l}(k^{\prime 2}-m_{c}^{2})^{m}[(y-x)^{2}]^{n}}.
  \end{equation}  
   The following identity,
   \begin{equation}
  	\frac{1}{[(y-x)^{2}]^{n}}=\int  \frac{\mathrm{d}^{D}t}{(2\pi)^{D}}e^{-i t.(y-x)}i(-1)^{n+1}2^{D-2n}\pi^{D/2}\frac{\Gamma(D/2-n)}{\Gamma(n)}(-\frac{1}{t^{2}})^{D/2-n},
  \end{equation}    
   simplifies the calculations through expressing a part of denominators in exponential form.The computations proceed with applying the $x_{\mu}\rightarrow i\frac{\partial}{\partial p_{\mu}}$ and $y_{\nu}\rightarrow -i\frac{\partial}{\partial p^{\prime}_{\nu}}$ transformations and using the definition of four-dimensional Dirac delta function to perform Fourier integrals and calculate the integrals over four $k$ and $k'$. The employment of Feynman parametrization and the  formula,
    \begin{equation}
   	\int \mathrm{d}^{D}t\frac{(t^2)^m}{(t^2+\Delta)^n}=\frac{i\pi^{2}(-1)^{m-n}\Gamma(m+2)\Gamma(n-m-\frac{D}{2})}{\Gamma(2)\Gamma(n)[-\Delta]^{n-m-\frac{D}{2}}},
   \end{equation}
    enable us to continue with more computations and perform the integrals. Subsequently, we utilize the following identity,
     \begin{equation}
    	\Gamma[\frac{D}{2}-n](-\frac{1}{\Delta})^{D/2-n}=\frac{(-1)^{n-1}}{(n-2)!}(-\Delta)^{n-2}~ln[-\Delta],
    \end{equation} 
     to obtain the imaginary parts of the ultimate results. Therefore, the QCD representation  of the correlation function, in respect of various Lorentz structures, is achieved as follows, 
      \begin{eqnarray}\label{QCD1}
     	&&\Pi_{\mu\nu}^{QCD}(p,p^{\prime},q^{2})=\Pi_{p_{\mu}p^{\prime}_{\nu}}^{QCD}(p^{2},p^{\prime2},q^{2})p_{\mu}p^{\prime}_{\nu}+\Pi_{p^{\prime}_{\mu}p^{\prime}_{\nu}}^{QCD}(p^{2},p^{\prime2},q^{2})p^{\prime}_{\mu}p^{\prime}_{\nu}+\Pi_{g_{\mu\nu}}^{QCD}(p^{2},p^{\prime2},q^{2})g_{\mu\nu}+\Pi_{p_{\mu}p^{\prime}_{\nu}\gamma_{5}}^{QCD}(p^{2},p^{\prime2},q^{2}) \,\nonumber\\ &&p_{\mu}p^{\prime}_{\nu}\gamma_{5}+\Pi_{p^{\prime}_{\mu}p^{\prime}_{\nu}\gamma_{5}}^{QCD}(p^{2},p^{\prime2},q^{2})p^{\prime}_{\mu}p^{\prime}_{\nu}\gamma_{5}+\Pi_{g_{\mu\nu}\gamma_{5}}^{QCD}(p^{2},p^{\prime2},q^{2})g_{\mu\nu}\gamma_{5}+\Pi_{p_{\mu}p^{\prime}_{\nu}\slashed{p}}^{QCD}(p^{2},p^{\prime2},q^{2}) p_{\mu}p^{\prime}_{\nu}\slashed{p}+\Pi_{p^{\prime}_{\mu}p^{\prime}_{\nu}\slashed{p}}^{QCD}(p^{2},p^{\prime2},q^{2})\nonumber\\&& p^{\prime}_{\mu}p^{\prime}_{\nu}\slashed{p}+\Pi_{g_{\mu\nu}\slashed{p}}^{QCD}(p^{2},p^{\prime2},q^{2}) g_{\mu\nu}\slashed{p}+\Pi_{p_{\mu}p^{\prime}_{\nu}\slashed{p}^{\prime}}^{QCD}(p^{2},p^{\prime2},q^{2}) p_{\mu}p^{\prime}_{\nu}\slashed{p}^{\prime}+\Pi_{p^{\prime}_{\mu}p^{\prime}_{\nu}\slashed{p}^{\prime}}^{QCD}(p^{2},p^{\prime2},q^{2}) p^{\prime}_{\mu}p^{\prime}_{\nu}\slashed{p}^{\prime}+\Pi_{g_{\mu\nu}\slashed{p}^{\prime}}^{QCD}(p^{2},p^{\prime2},q^{2})\nonumber\\&& g_{\mu\nu}\slashed{p}^{\prime}+\Pi_{p^{\prime}_{\mu}p^{\prime}_{\nu}\slashed{p}}^{QCD}(p^{2},p^{\prime2},q^{2}) p^{\prime}_{\mu}p^{\prime}_{\nu}\slashed{p}+\Pi_{p_{\mu}p^{\prime}_{\nu}\slashed{p}\gamma_{5}}^{QCD}(p^{2},p^{\prime2},q^{2}) p_{\mu}p^{\prime}_{\nu}\slashed{p}\gamma_{5}+\Pi_{p^{\prime}_{\mu}p^{\prime}_{\nu}\slashed{p}\gamma_{5}}^{QCD}(p^{2},p^{\prime2},q^{2}) p^{\prime}_{\mu}p^{\prime}_{\nu}\slashed{p}\gamma_{5}+\nonumber\\&&\Pi_{g_{\mu\nu}\slashed{p}\gamma_{5}}^{QCD}(p^{2},p^{\prime2},q^{2}) g_{\mu\nu}\slashed{p}\gamma_{5}+\Pi_{p_{\mu}p^{\prime}_{\nu}\slashed{p}\slashed{p}^{\prime}}^{QCD}(p^{2},p^{\prime2},q^{2}) p_{\mu}p^{\prime}_{\nu}\slashed{p}\slashed{p}^{\prime}+\Pi_{p^{\prime}_{\mu}p^{\prime}_{\nu}\slashed{p}\slashed{p}^{\prime}}^{QCD}(p^{2},p^{\prime2},q^{2}) p^{\prime}_{\mu}p^{\prime}_{\nu}\slashed{p}\slashed{p}^{\prime}+\Pi_{g_{\mu\nu}\slashed{p}\slashed{p}^{\prime}}^{QCD}(p^{2},p^{\prime2},q^{2})\nonumber\\&& g_{\mu\nu}\slashed{p}\slashed{p}^{\prime}+\Pi_{p_{\mu}p^{\prime}_{\nu}\slashed{p}^{\prime}\gamma_{5}}^{QCD}(p^{2},p^{\prime2},q^{2}) p_{\mu}p^{\prime}_{\nu}\slashed{p}^{\prime}\gamma_{5}+\Pi_{p^{\prime}_{\mu}p^{\prime}_{\nu}\slashed{p}^{\prime}\gamma_{5}}^{QCD}(p^{2},p^{\prime2},q^{2}) p^{\prime}_{\mu}p^{\prime}_{\nu}\slashed{p}^{\prime}\gamma_{5}+\Pi_{g_{\mu\nu}\slashed{p}^{\prime}\gamma_{5}}^{QCD}(p^{2},p^{\prime2},q^{2}) g_{\mu\nu}\slashed{p}^{\prime}\gamma_{5}+\nonumber\\&&\Pi_{p^{\prime}_{\mu}p^{\prime}_{\nu}\slashed{p}\gamma_{5}}^{QCD}(p^{2},p^{\prime2},q^{2}) p^{\prime}_{\mu}p^{\prime}_{\nu}\slashed{p}\gamma_{5}+\Pi_{p^{\prime}_{\nu}\gamma_{\mu}\slashed{p}}^{QCD}(p^{2},p^{\prime2},q^{2}) p^{\prime}_{\nu}\gamma_{\mu}\slashed{p}+\Pi_{p^{\prime}_{\nu}\gamma_{\mu}\slashed{p}^{\prime}}^{QCD}(p^{2},p^{\prime2},q^{2}) p^{\prime}_{\nu}\gamma_{\mu}\slashed{p}^{\prime}+\nonumber\\&&\Pi_{p_{\mu}p^{\prime}_{\nu}\slashed{p}\slashed{p}^{\prime}\gamma_{5}}^{QCD}(p^{2},p^{\prime2},q^{2}) p_{\mu}p^{\prime}_{\nu}\slashed{p}\slashed{p}^{\prime}\gamma_{5} +\Pi_{p^{\prime}_{\mu}p^{\prime}_{\nu}\slashed{p}\slashed{p}^{\prime}\gamma_{5}}^{QCD}(p^{2},p^{\prime2},q^{2}) p^{\prime}_{\mu}p^{\prime}_{\nu}\slashed{p}\slashed{p}^{\prime}\gamma_{5}+\Pi_{g_{\mu\nu}\slashed{p}\slashed{p}^{\prime}\gamma_{5}}^{QCD}(p^{2},p^{\prime2},q^{2}) \,g_{\mu\nu}\slashed{p}\slashed{p}^{\prime}\gamma_{5}+\nonumber\\&&\Pi_{p^{\prime}_{\nu}\gamma_{\mu}\slashed{p}\gamma_{5}}^{QCD}(p^{2},p^{\prime2},q^{2}) p^{\prime}_{\nu}\gamma_{\mu}\slashed{p}\gamma_{5}+\Pi_{p^{\prime}_{\nu}\gamma_{\mu}\slashed{p}\slashed{p}^{\prime}}^{QCD}(p^{2},p^{\prime2},q^{2}) p^{\prime}_{\nu}\gamma_{\mu}\slashed{p}\slashed{p}^{\prime}+\Pi_{p^{\prime}_{\nu}\gamma_{\mu}\slashed{p}^{\prime}\gamma_{5}}^{QCD}(p^{2},p^{\prime2},q^{2}) p^{\prime}_{\nu}\gamma_{\mu}\slashed{p}^{\prime}\gamma_{5}+\nonumber\\&&\Pi_{p^{\prime}_{\nu}\gamma_{\mu}\slashed{p}\slashed{p}^{\prime}\gamma_{5}}^{QCD}(p^{2},p^{\prime2},q^{2}) p^{\prime}_{\nu}\gamma_{\mu}\slashed{p}\slashed{p}^{\prime}\gamma_{5},
     \end{eqnarray}
     in which the invariant functions, $\Pi_{i}^{QCD}(p^{2},p^{\prime2},q^{2})$($i$ stands for individual Lorentz structure), as the coefficients of the structures, are characterized in terms of the double dispersion integrals which arise from the contributions with imaginary parts, $\rho_{i}^{QCD}(s,s^{\prime},q^{2})=\frac{1}{\pi}Im\Pi_{i}^{QCD}(p^{2},p^{\prime2},q^{2})$, and $\Gamma_{i}(p^{2},p^{\prime2},q^{2})$, representing the contributions without imaginary parts as,
      \begin{equation}\label{spectraldensity}
     	\Pi_{i}^{QCD}(p^{2},p^{\prime2},q^{2})=\int_{s_{min}}^{\infty}\mathrm{d}s\int_{s^{\prime}_{min}}^{\infty}\mathrm{d}s^{\prime}\frac{\rho_{i}^{QCD}(s,s^{\prime},q^{2})}{(s-p^{2})(s^{\prime}-p^{\prime2})}+\Gamma_{i}(p^{2},p^{\prime2},q^{2}).
     \end{equation} 
    where  $s_{min}=(m_{b}+m_s+m_s)^{2}$, $s^{\prime}_{min}=(m_{c}+m_s+m_s)^{2}$. The spectral density, $\rho_{i}^{QCD}(s,s^{\prime},q^{2})$, incorporates both the perturbative, $\rho_{i}^{Pert.}(s,s^{\prime},q^{2})$, and nonperturbative, $\sum_{n=3}^{6}\rho_{i}^{n}(s,s^{\prime},q^{2})$, contributions which correspond to the vacuum condensates of different dimensions,   
     \begin{equation}
    	\rho_{i}^{QCD}(s,s^{\prime},q^{2})=\rho_{i}^{Pert.}(s,s^{\prime},q^{2})+\sum_{n=3}^{6}\rho_{i}^{n}(s,s^{\prime},q^{2}).	 
    \end{equation} 
   Regarding the quark-hadron duality assumption, the integrals' upper bounds in Eq. \eqref{spectraldensity} change to $s_0$ and $s'_0$ as continuum thresholds of the initial and final baryonic states, respectively. In order to suppress the contributions of higher states and continuum on the QCD side, in a similar way to the physical side of the correlation function, it is required to employ the double Borel transformation, Eq. \eqref{Borel transformation}, and continuum subtraction, originated from the quark-hadron duality supposition. Hence, the invariant functions receive the following form,
    \begin{equation}
   	\Pi_{i}^{QCD}(M^{2},M^{\prime2},s_{0},s^{\prime}_{0},q^{2})=\int_{s_{min}}^{s_{0}}\mathrm{d}s\int_{s^{\prime}_{min}}^{s^{\prime}_{0}}\mathrm{d}s^{\prime}e^{-s/M^{2}}e^{-s^{\prime}/M^{\prime2}}\rho_{i}^{QCD}(s,s^{\prime},q^{2})	+\mathbf{\widehat{B}}\big[\Gamma_{i}(p^{2},p^{\prime2},q^{2})\big],
   \end{equation}  
    The definite expressions of the spectral densities, $\rho_{i}(s,s^{\prime},q^{2})$, and $\Gamma_{i}(p^2,p^{\prime2},q^2)$ functions associated with the structure $g_{\mu\nu}\slashed {p}'\gamma_5$ are presented in appendix \ref{AppC}. Subsequently, as a final step, we match the corresponding coefficients of the Lorentz structures from the phenomenological and QCD sides to derive the sum rules of the transition form factors in respect of hadronic and QCD parameters, such as the masses and residues of the initial and final baryons, quark masses, and vacuum condensates including quark, gluon, and mixed condensates; as well as auxiliary parameters $s_{0}, s^{\prime}_{0}, M^{2}, M^{\prime2}$, and $\beta$. We provide the representations of the sum rules associated with the transition form factors in appendix \ref{AppB}.  
    \section{NUMERICAL ANALYSIS}\label{numerical}
 Form factors are considered as the main features to parametrize  hadronic transitions and provide an extensive understanding of the hadrons' properties and structures. In this context, form factors  are provided as significant input into the survey of the observable quantities such as decay width and branching fraction in various decay channels. As previously mentioned, the form factors are established by relating the corresponding coefficients of the structures from both the phenomenological and QCD sides and obtaining the appropriate sum rules, demanded of the method, in respect of the QCD, hadronic, and auxiliary parameters. These fundamental quantities are Lorentz invariant with $q^2$ dependency. Consequently, it is required to analyze their properties in respect of $q^2$. Before proceeding with numerical analysis of these form factors, it is essential to evaluate the working intervals of the auxiliary parameters, involved in this approach, including Borel parameters $M^2$ and $M'^2$, continuum thresholds $s_0$ and $s'_0$, and the mixing parameter $\beta$, regarding the basic principles of this method.  To this end,  we need some input parameters expressed in Table \ref{inputparameters}.
 
 The physical quantities are generally supposed to remain steady in respect of the fluctuations in these parameters. Additionally, pole dominance and convergence of the OPE, as the main considerations of the QCD sum rules method, effectively restrict the working areas of these helping parameters. Pole dominance requires that the ground state's contribution must be dominant in comparison to the contributions of higher modes and continuum. Hence, the upper bounds of the Borel parameters $M^2$ and $M'^2$ are determined through demanding the following condition:
    \begin{equation}
    	PC=\frac{\Pi^{QCD}(M^{2},M^{\prime2},s_{0},s^{\prime}_{0})}{\Pi^{QCD}(M^{2},M^{\prime2},\infty,\infty)}\ge\frac{1}{2}.
    \end{equation}
    The OPE series incorporates both perturbative and nonperturbative contributions. The convergence of this series is satisfied with the more contribution of the perturbative part rather than the nonperturbative sectors. Furthermore, among the nonperturbative parts, those with higher dimensions receive lower contributions that it confirms the convergence of the OPE and evaluates the lower limits of the Borel parameters. To this purpose, the following constraint is applied,
     \begin{equation}
    	R(M^{2},M^{\prime2})=\frac{\Pi^{QCD-dim6}(M^{2},M^{\prime2},s_{0},s^{\prime}_{0})}{\Pi^{QCD}(M^{2},M^{\prime2},s_{0},s_{0}^{\prime})}\le~0.05.
    \end{equation}   
    Taking these conditions into account, the proper intervals of the Borel parameters for the initial and final states are respectively achieved as,
    \begin{eqnarray}
    && 9\mathrm{GeV^{2}}\le~M^{2}\le~12\mathrm{GeV^{2}},\nonumber\\
    && 6\mathrm{GeV^{2}}\le~M^{\prime2}~\le~9\mathrm{GeV^{2}}.
    \end{eqnarray}

    The other auxiliary parameters, which restrict the upper bounds of the integrals in computations, are continuum thresholds, $s_{0}$ and $s'_{0}$. These parameters arise from the quark-hadron duality assumption that is employed to eliminate any contribution to the excited modes and continuum in the initial and final states. In a similar way to the Borel parameters, the ultimate results of the sum rules are considered to be stable within the practical intervals of the continuum threshold parameters. Ensuring this stability, we select,
    \begin{eqnarray}
  	&&(m_{\Omega_{b}^{*}}+0.1)^{2}~\mathrm{GeV^{2}}\le s_{0}\le(m_{\Omega_{b}^{*}}+0.5)^{2}~\mathrm{GeV^{2}},\notag\\ 
   	&&(m_{\Omega_{c}}+0.1)^{2}~\mathrm{GeV^{2}}\le s_{0}^{\prime}\le(m_{\Omega_{c}}+0.5)^{2}~\mathrm{GeV^{2}}.	
    \end{eqnarray}

\begin{figure}
    	\begin{center}
    	\includegraphics[height=5cm,width=5.5cm]{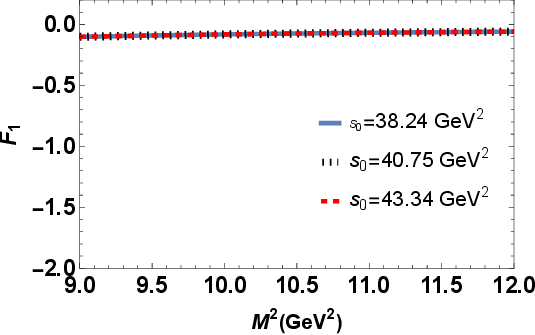}
    	\includegraphics[height=5cm,width=5.5cm]{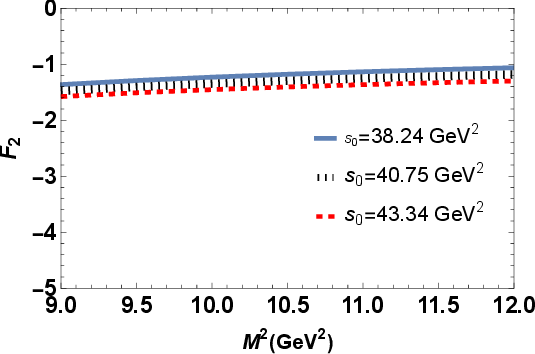}
    	\includegraphics[height=5cm,width=5.5cm]{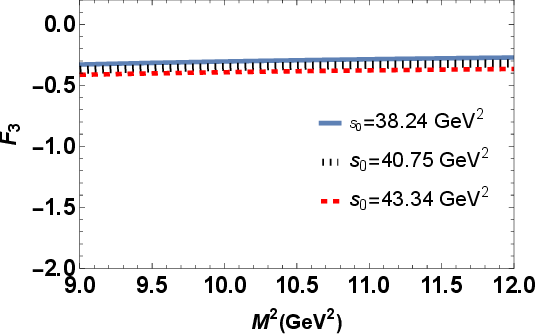}
    	\includegraphics[height=5cm,width=5.5cm]{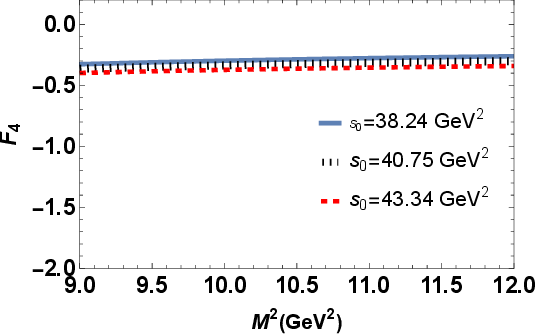}
    	\includegraphics[height=5cm,width=5.5cm]{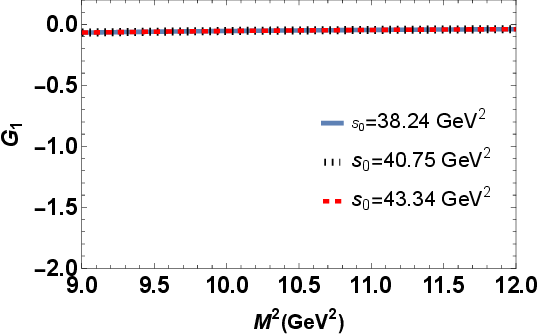}
    	\includegraphics[height=5cm,width=5.5cm]{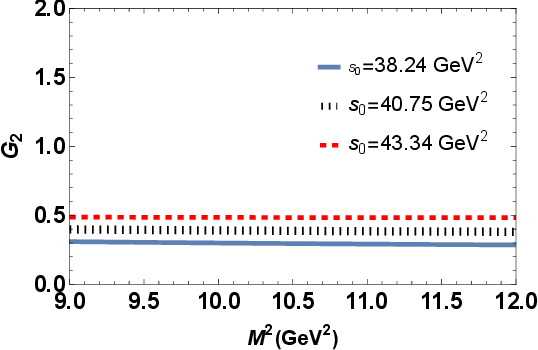}
    	\includegraphics[height=5cm,width=5.5cm]{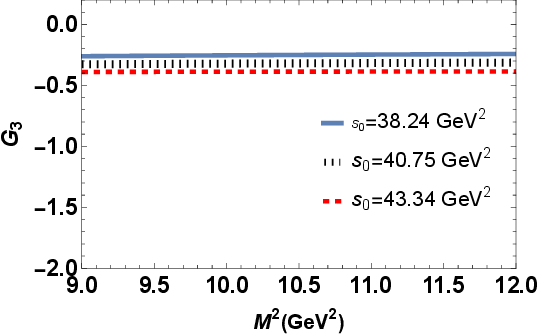}
    	\includegraphics[height=5cm,width=5.5cm]{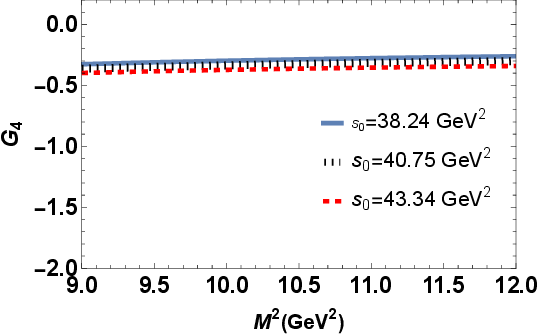}
    	\end{center}
    	\caption{Behavior of the form factors in relation to the Borel parameter $M^2$ for different values of the parameter $s_0$, $q^{2}=0$, and the central values of the other helping parameters. The graphs are associated with the structures $g_{\mu\nu}\slashed{p}^{\prime}\gamma_5, p^{\prime}_{\nu}\gamma_{\mu}\slashed{p}\slashed{p}^{\prime}\gamma_{5}, p_{\mu}p^{\prime}_{\nu}\slashed{p}^{\prime}\gamma_{5}, p^{\prime}_{\mu}p^{\prime}_{\nu}\slashed{p}\slashed{p}^{\prime}\gamma_{5}, g_{\mu\nu}\slashed{p}, p^{\prime}_{\nu}\gamma_{\mu}\slashed{p}^{\prime}, p_{\mu}p^{\prime}_{\nu}\slashed{p},$ and $p^{\prime}_{\mu}p^{\prime}_{\nu}\slashed{p}\slashed{p}^{\prime}$ matching to the form factors $F_{1}, F_{2}, F_{3}, F_{4}, G_{1}, G_{2}, G_{3}$, and $G_{4}$, respectively, of the $\Omega_{b}^{*}\rightarrow \Omega_{c}\ell\bar{\nu}_{\ell}$ transition, 
    	 (see Table \ref{fitparameters1}).} 
    	\label{ffM2s0}
    \end{figure}

\begin{figure}
    	\begin{center}
    		\includegraphics[height=5cm,width=5.5cm]{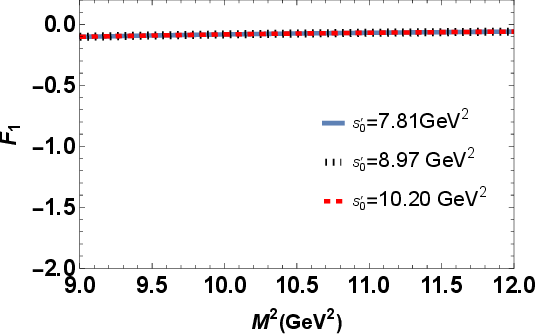}
    		\includegraphics[height=5cm,width=5.5cm]{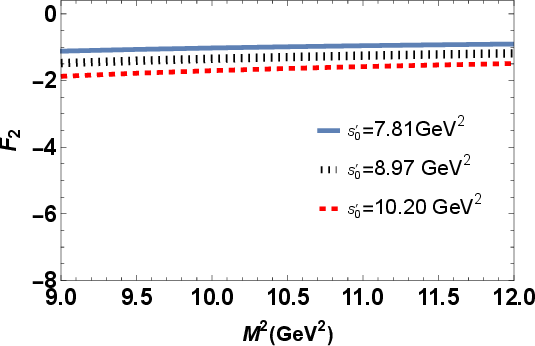}
    		\includegraphics[height=5cm,width=5.5cm]{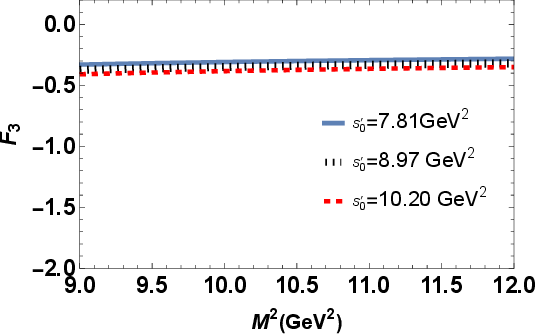}
    		\includegraphics[height=5cm,width=5.5cm]{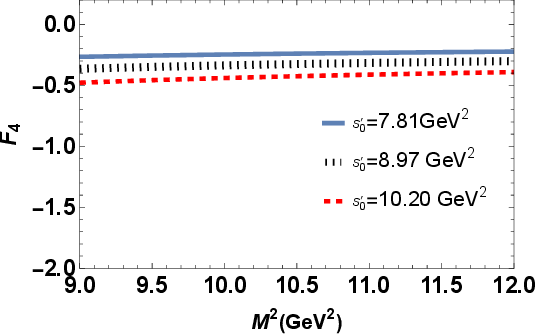}
    		\includegraphics[height=5cm,width=5.5cm]{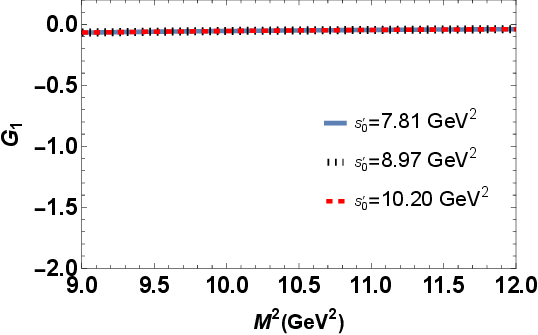}
    		\includegraphics[height=5cm,width=5.5cm]{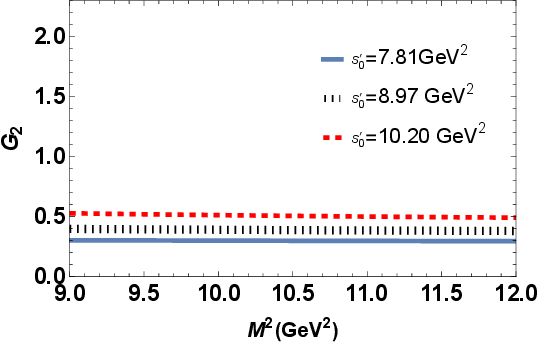}
    		\includegraphics[height=5cm,width=5.5cm]{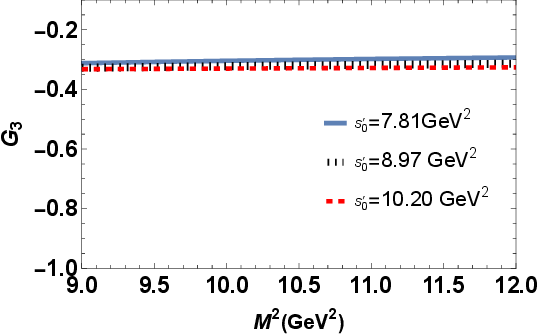}
    		\includegraphics[height=5cm,width=5.5cm]{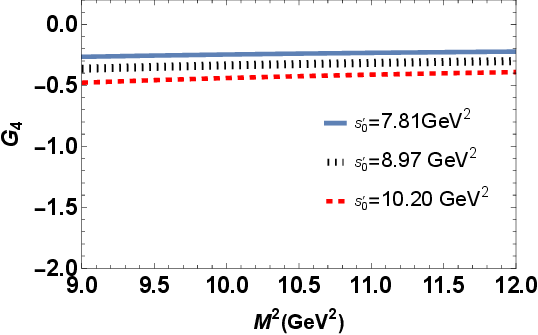}
    	\end{center}
    	\caption{Behavior of the form factors in relation to the Borel parameter $M^2$ for different values of the parameter $s'_0$, $q^{2}=0$, and the central values of the other helping parameters. The graphs are associated with the structures $g_{\mu\nu}\slashed{p}^{\prime}\gamma_5, p^{\prime}_{\nu}\gamma_{\mu}\slashed{p}\slashed{p}^{\prime}\gamma_{5}, p_{\mu}p^{\prime}_{\nu}\slashed{p}^{\prime}\gamma_{5}, p^{\prime}_{\mu}p^{\prime}_{\nu}\slashed{p}\slashed{p}^{\prime}\gamma_{5}, g_{\mu\nu}\slashed{p}, p^{\prime}_{\nu}\gamma_{\mu}\slashed{p}^{\prime}, p_{\mu}p^{\prime}_{\nu}\slashed{p},$ and $p^{\prime}_{\mu}p^{\prime}_{\nu}\slashed{p}\slashed{p}^{\prime}$ matching to the form factors $F_{1}, F_{2}, F_{3}, F_{4}, G_{1}, G_{2}, G_{3}$, and $G_{4}$, respectively,  
        of the $\Omega_{b}^{*}\rightarrow \Omega_{c}\ell\bar{\nu}_{\ell}$ transition, 
    	(see Table \ref{fitparameters1}).} 
    	\label{ffM2s0p}
    \end{figure}

\begin{figure}
    	\begin{center}
    		\includegraphics[height=5cm,width=5.5cm]{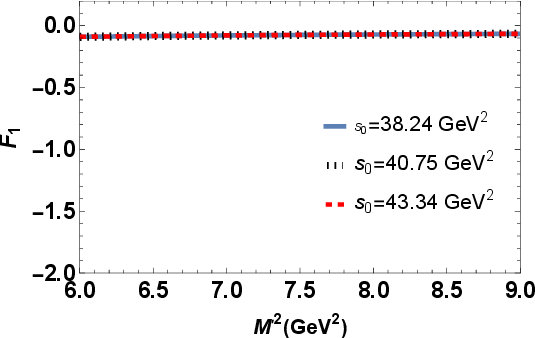}
    		\includegraphics[height=5cm,width=5.5cm]{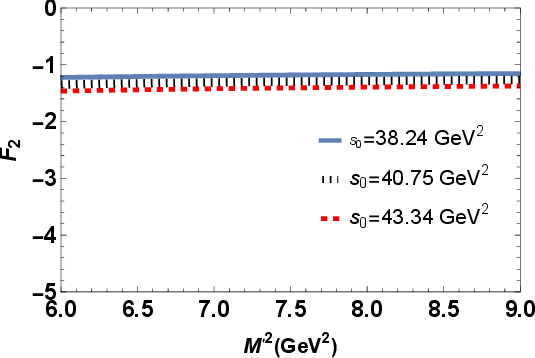}
    		\includegraphics[height=5cm,width=5.5cm]{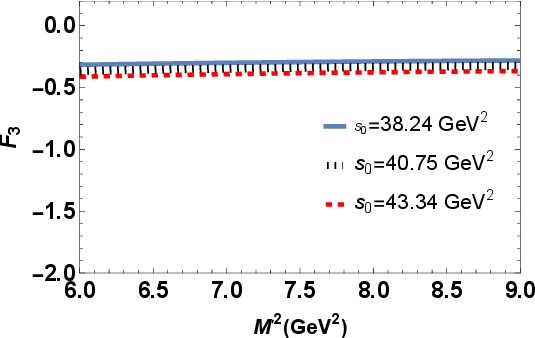}
    		\includegraphics[height=5cm,width=5.5cm]{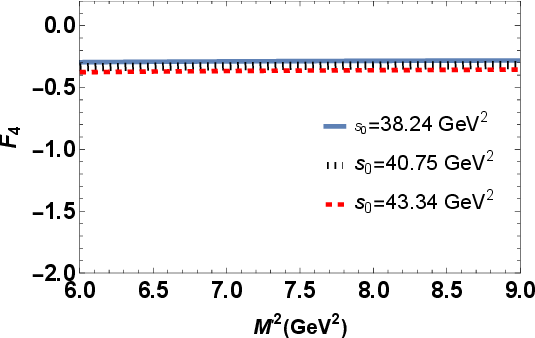}
    		\includegraphics[height=5cm,width=5.5cm]{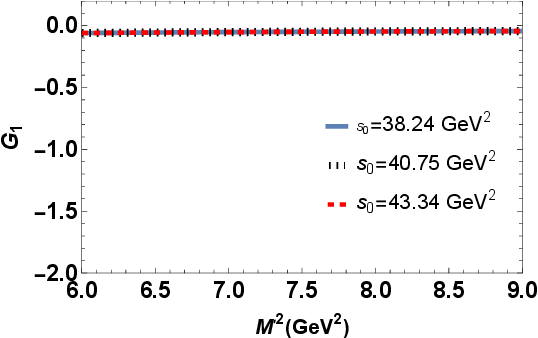}
    		\includegraphics[height=5cm,width=5.5cm]{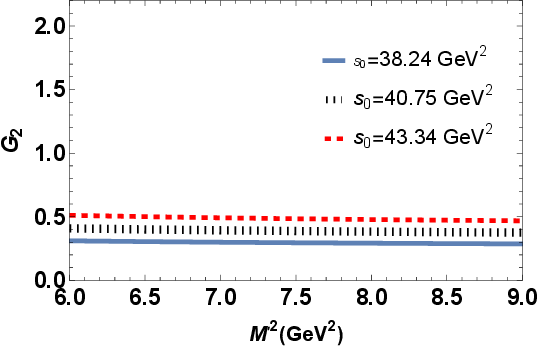}
    		\includegraphics[height=5cm,width=5.5cm]{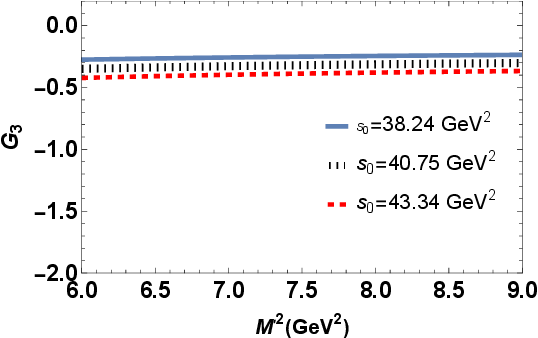}
    		\includegraphics[height=5cm,width=5.5cm]{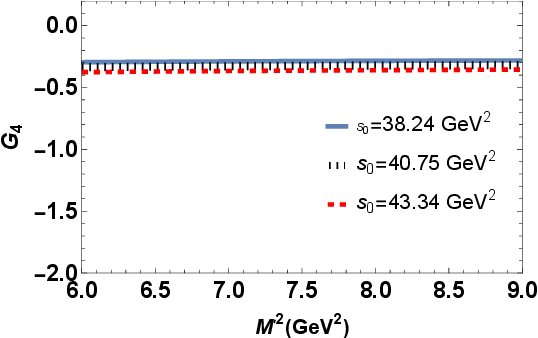}
    	\end{center}
    	\caption{Behavior of the form factors in relation to the Borel parameter $M'^2$ for different values of the parameter $s_0$, $q^{2}=0$, and the central values of the other helping parameters. The graphs are associated with the structures $g_{\mu\nu}\slashed{p}^{\prime}\gamma_5, p^{\prime}_{\nu}\gamma_{\mu}\slashed{p}\slashed{p}^{\prime}\gamma_{5}, p_{\mu}p^{\prime}_{\nu}\slashed{p}^{\prime}\gamma_{5}, p^{\prime}_{\mu}p^{\prime}_{\nu}\slashed{p}\slashed{p}^{\prime}\gamma_{5}, g_{\mu\nu}\slashed{p}, p^{\prime}_{\nu}\gamma_{\mu}\slashed{p}^{\prime}, p_{\mu}p^{\prime}_{\nu}\slashed{p},$ and $p^{\prime}_{\mu}p^{\prime}_{\nu}\slashed{p}\slashed{p}^{\prime}$ matching to the form factors $F_{1}, F_{2}, F_{3}, F_{4}, G_{1}, G_{2}, G_{3}$, and $G_{4}$, respectively, of the $\Omega_{b}^{*}\rightarrow \Omega_{c}\ell\bar{\nu}_{\ell}$ transition, 
    		(see Table \ref{fitparameters1}) 
    		.} 
    	\label{ffMp2s0}
    \end{figure}

\begin{figure}
    	\begin{center}
    		\includegraphics[height=5cm,width=5.5cm]{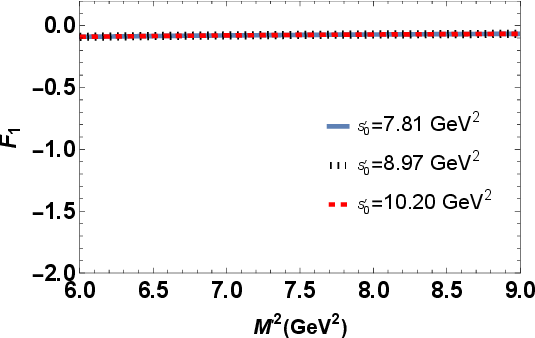}
    		\includegraphics[height=5cm,width=5.5cm]{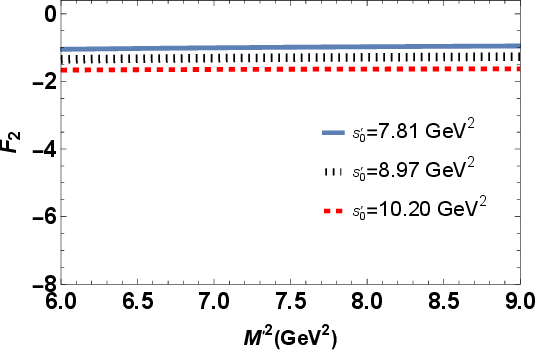}
    		\includegraphics[height=5cm,width=5.5cm]{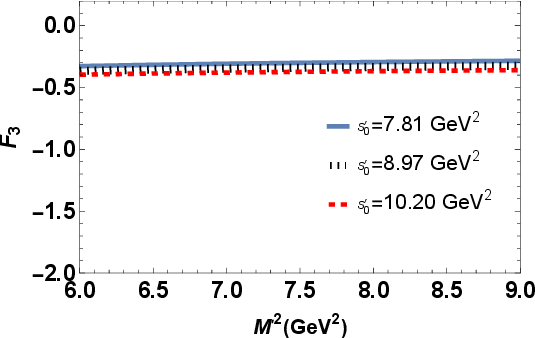}
    		\includegraphics[height=5cm,width=5.5cm]{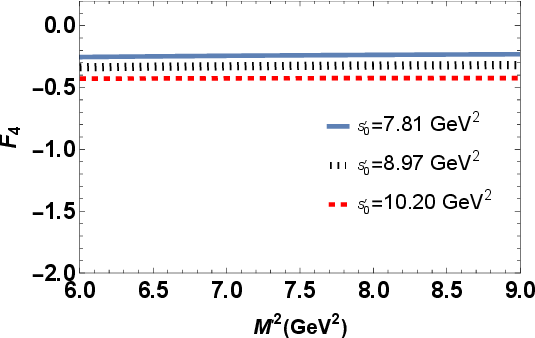}
    		\includegraphics[height=5cm,width=5.5cm]{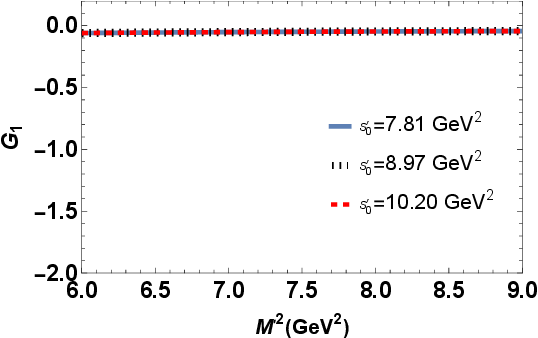}
    		\includegraphics[height=5cm,width=5.5cm]{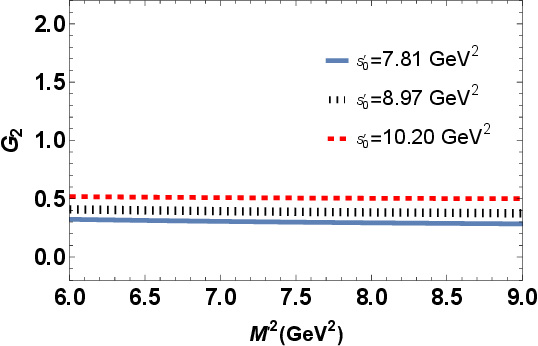}
    		\includegraphics[height=5cm,width=5.5cm]{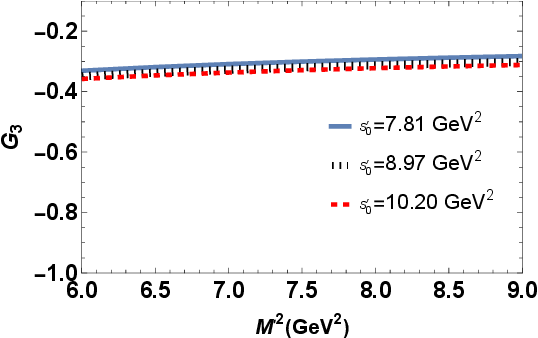}
    		\includegraphics[height=5cm,width=5.5cm]{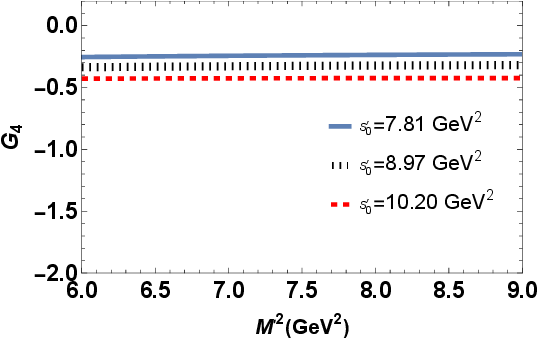}
    	\end{center}
    	\caption{Behavior of the form factors in relation to the Borel parameter $M'^2$ for different values of the parameter $s'_0$, $q^{2}=0$, and the central values of the other helping parameters. The graphs are associated with the structures $g_{\mu\nu}\slashed{p}^{\prime}\gamma_5, p^{\prime}_{\nu}\gamma_{\mu}\slashed{p}\slashed{p}^{\prime}\gamma_{5}, p_{\mu}p^{\prime}_{\nu}\slashed{p}^{\prime}\gamma_{5}, p^{\prime}_{\mu}p^{\prime}_{\nu}\slashed{p}\slashed{p}^{\prime}\gamma_{5}, g_{\mu\nu}\slashed{p}, p^{\prime}_{\nu}\gamma_{\mu}\slashed{p}^{\prime}, p_{\mu}p^{\prime}_{\nu}\slashed{p},$ and $p^{\prime}_{\mu}p^{\prime}_{\nu}\slashed{p}\slashed{p}^{\prime}$ matching to the form factors $F_{1}, F_{2}, F_{3}, F_{4}, G_{1}, G_{2}, G_{3}$, and $G_{4}$, respectively of the $\Omega_{b}^{*}\rightarrow \Omega_{c}\ell\bar{\nu}_{\ell}$ transition, 
    		(see Table \ref{fitparameters1}).} 
    	\label{ffMp2s0p}
    \end{figure}

    The consistency between the requirements of this approach and the selected ranges of the parameters is illustrated in Figs. \ref{ffM2s0}, \ref{ffM2s0p}, \ref{ffMp2s0}, and \ref{ffMp2s0p} in which the form factors indicate an acceptable steadiness in respect of variations of the parameters, $M^2$, $M'^2$, $s_0$, and $s'_0$ within the allowed regions.
     \begin{figure}
    	\begin{center}
    		\includegraphics[height=5cm,width=5.5cm]{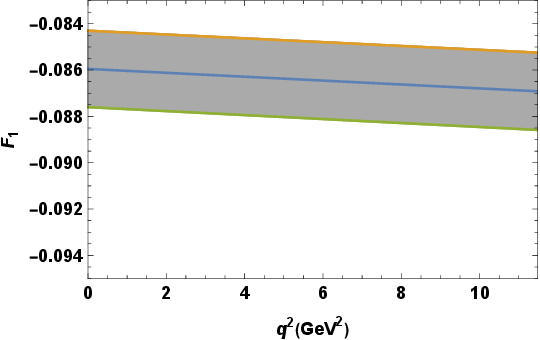}
    		\includegraphics[height=5cm,width=5.5cm]{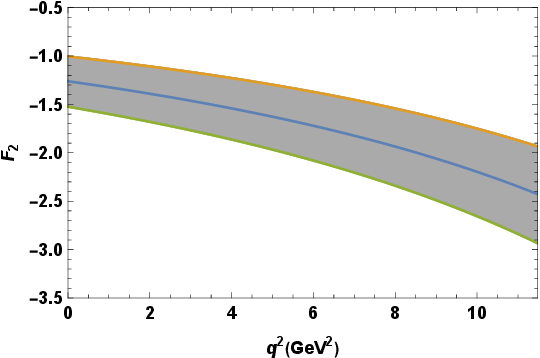}
    		\includegraphics[height=5cm,width=5.5cm]{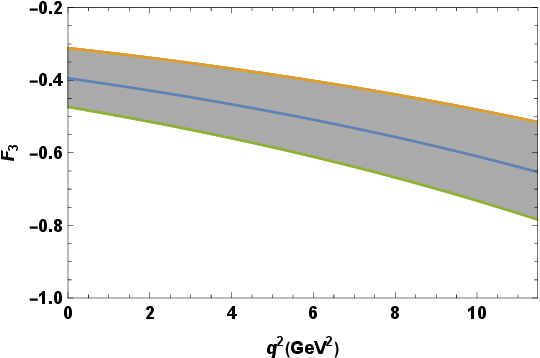}
    		\includegraphics[height=5cm,width=5.5cm]{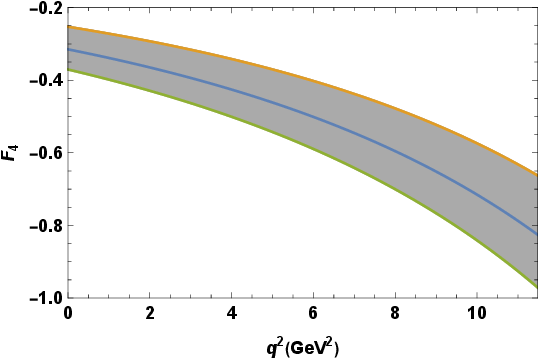}
    		\includegraphics[height=5cm,width=5.5cm]{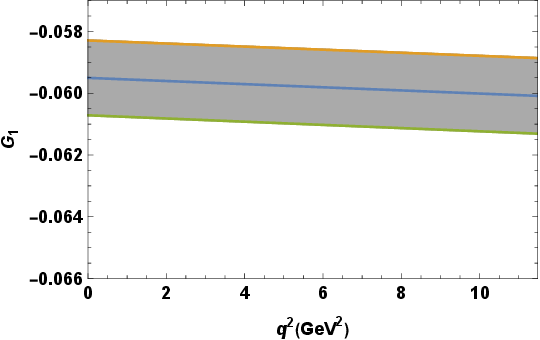}
    		\includegraphics[height=5cm,width=5.5cm]{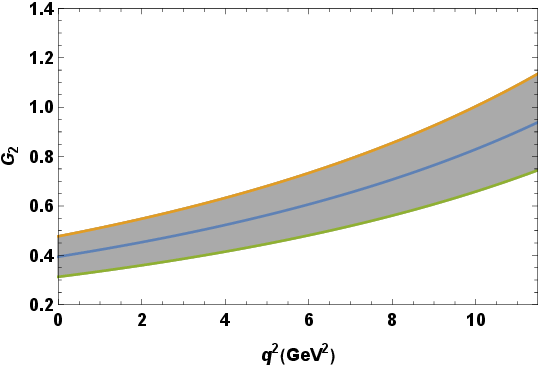}
    		\includegraphics[height=5cm,width=5.5cm]{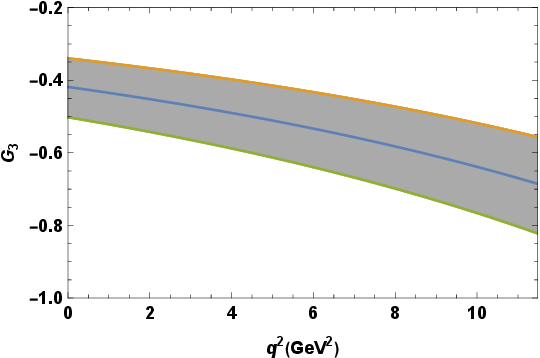}
    		\includegraphics[height=5cm,width=5.5cm]{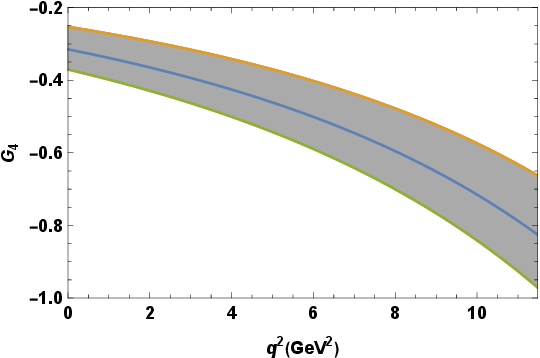}
    	\end{center}
    	\caption{Behavior of the form factors $F_{1}, F_{2}, F_{3}, F_{4}, G_{1}, G_{2}, G_{3}$, and $G_{4}$ of the $\Omega_{b}^{*}\rightarrow \Omega_{c}\ell\bar{\nu}_{\ell}$ transition, in relation to $q^2$ for central values of the helping parameters, considering the error bands, presented in Table \ref{fitparameters1}.} 
    	\label{ffq2error}
    \end{figure}
   An arbitrary linear combination in the interpolating currents of spin-$\frac{1}{2}$ baryons introduces a mathematical mixing parameter, $\beta$, in this approach, that is  essential to be identified for analyzing the form factors and the other physical quantities. This parameter is valid within an extensive range from $-\infty$ to $+\infty$ and we utilize the definition $x=\cos\theta$ where $\theta=\tan^{-1}\beta$. To satisfy the QCD sum rules' requirements, the mixing parameter $\beta$, like the other auxiliary parameters, is determined in a manner that guarantees the stability of the form factors within the selected interval. Hence, the operative region of the parameter $x$ is obtained as $-1.0\le x\le-0.5$ that all responsible form factors indicate an acceptable stability and slight dependence on variations within this interval which includes $x=-0.71$ related to $\beta=-1$ in the Ioffe current and it is employed as an arbitrary point in numerical analysis.
   
   As mentioned earlier, the form factors are $q^2$ dependent. Therefore, the next step, after determination of the practical areas of the helping parameters, is to investigate the variations of the form factors in relation to $q^2$ in whole physical region, $m_{\ell}^{2}\le~q^{2}~\le (m_{\Omega_{b}^{*}}-m_{\Omega_{c}})^{2}$, in which the maximum value of $q^2$ is $q^2_{max}=11.48~\mathrm{GeV^2}$ and the lowest value is related to the squared lepton's mass, itemized in Table \ref{inputparameters}.
     \begin{table}[h!]
    	\centering
    	\caption{ Input QCD and hadronic parameters utilized in numerical analysis.}
    	\begin{tabular}{cc|cc}
    		\hline
    		Parameter&Value&~Parameter&Value\\
    		\hline
    		$m_{b}$&$(4.18^{+0.03}_{-0.02})~\mathrm{GeV}$\cite{ParticleDataGroup:2024cfk}&$m_{\Omega_{b}^{*}}$& $(6084\pm84)~\mathrm{MeV}$\cite{Agaev:2017jyt}\\
    		$m_{c}$&$(1.27\pm0.02)~\mathrm{GeV}$\cite{ParticleDataGroup:2024cfk}&	$m_{\Omega_{c}}$&$(2695.3\pm0.4)~\mathrm{MeV}$\cite{ParticleDataGroup:2024cfk}\\
    		$m_{e}$&$0.51~\mathrm{MeV}$\cite{ParticleDataGroup:2024cfk}&	$\lambda_{\Omega_{b}^{*}}$&$(0.093\pm0.014)~\mathrm{GeV^{3}}$\cite{Agaev:2017jyt}\\
    		$m_{\mu}$&$105.65~\mathrm{MeV}$\cite{ParticleDataGroup:2024cfk}&	$\lambda_{\Omega_{c}}$&$(0.062\pm0.018)~\mathrm{GeV^{3}}$\cite{Agaev:2017jyt} \\
    		$m_{\tau}$&$1776.93~\mathrm{MeV}$\cite{ParticleDataGroup:2024cfk} &$m_{\Omega_{b}}$& $(6045.8\pm0.8)~\mathrm{MeV}$\cite{ParticleDataGroup:2024cfk}\\
    		$G_{F}$&$1.17\times 10^{-5}~\mathrm{GeV^{-2}}$\cite{ParticleDataGroup:2024cfk}&$m_{\Omega_{c}^{*}}$&$(2765.9\pm2.0)~\mathrm{MeV}$\cite{ParticleDataGroup:2024cfk}\\
    		$V_{cb}$&$(40\pm 1.1)\times 10^{-3}$\cite{ParticleDataGroup:2024cfk}&$m_{s}$&$(93.5\pm0.8)~\mathrm{MeV}$\cite{ParticleDataGroup:2024cfk}\\
    		$m_{0}^{2}$&$(0.8 \pm 0.2)~\mathrm{GeV^{2}}$\cite{Belyaev:1982cd,Belyaev:1982sa,Ioffe:2005ym}&$\lambda_{\Omega_{c}^{*}}$&$(0.071\pm0.010)~\mathrm{GeV^{3}}$\cite{Agaev:2017jyt}\\
    		$\langle \bar{u}u \rangle$&$-(0.24\pm 0.01)^{3}~\mathrm{GeV^{3}}$\cite{Belyaev:1982cd,Belyaev:1982sa}&$\langle \bar{s}s \rangle$&$(0.8\pm0.1)\langle \bar{u}u \rangle~\mathrm{GeV^3}$\cite{Belyaev:1982cd,Belyaev:1982sa}\\
    		$\langle 0|\frac{1}{\pi}\alpha_{s}~\mathrm{G^{2}}|0 \rangle$&$(0.012\pm 0.004)~\mathrm{GeV^{4}}$\cite{Belyaev:1982cd,Belyaev:1982sa,Ioffe:2005ym}&$\lambda_{\Omega_{b}}$&$(0.134\pm0.030)~\mathrm{GeV^{3}}$\cite{Wang:2009cr}\\
    		\hline
    		\hline
    	\end{tabular}
    	\label{inputparameters}
    \end{table}
     Our investigation reveals that the form factors are properly fitted to the following function,
    \begin{equation}
    	\mathcal{F}(q^{2})	= \frac{\mathcal{F}(0)}{\bigg(1-a_{1}\frac{q^{2}}{m_{\Omega_{b}^{*}}^{2}}+a_{2}\frac{q^{4}}{m_{\Omega_{b}^{*}}^{4}}+\,a_{3}\frac{q^{6}}{m_{\Omega_{b}^{*}}^{6}}+a_{4}\frac{q^{8}}{m_{\Omega_{b}^{*}}^{8}}\bigg)}\cdot
    \end{equation}

   At the central values of the helping parameters and Ioffe point, $x=-0.71$, the values of the fit parameters $\mathcal{F}(0), a_1, a_2, a_3, $ and $ a_4$, for the $\Omega^{*}_b\rightarrow\Omega_c\ell\bar{\nu}_{\ell}$ transition, are determined as presented in Table \ref{fitparameters1}. The findings related to the other transition, $\Omega_{b}\rightarrow \Omega_{c}^{*}\ell\bar{\nu}_{\ell}$, are displayed in Table \ref{fitparameters2}.  
    \begin{table}
   	\centering
   	\caption{Fit function parameters for the form factors of the $\Omega_{b}^{*}\rightarrow \Omega_{c}\ell\bar{\nu}_{\ell}$ semileptonic weak transition, considering uncertainties from both auxiliary and input QCD parameters.}
   	\begin{ruledtabular}
   		\begin{tabular}{|c|c|c|c|c|c|c|c|c|}
   			&$F_{1}(q^{2})$&$F_{2}(q^{2})$&$F_{3}(q^{2})$&$F_{4}(q^{2})$&$G_{1}(q^{2})$&$G_{2}(q^{2})$&$G_{3}(q^{2})$&$G_{4}(q^{2})$\\
   			\hline
   			$\mathcal{F}(q^{2}=0)$&$-0.085\pm0.003$&$-1.26\pm0.34$&$-0.39\pm0.11$&$-0.31\pm0.08$&$-0.059\pm0.002$&$0.39\pm0.11$&$-0.41\pm0.11$&$-0.31\pm0.08$\\
   			$a_{1}$&0.036&1.75&1.50&2.65&0.031&2.53&1.40&2.65\\
   			$a_{2}$&0.0013&0.61&0.74&2.33&0.001&2.59&0.43&2.33\\
   			$a_{3}$&-0.00005&0.13&-0.08&-0.68&-0.00003&-1.59&0.08&-0.68\\
   			$a_{4}$&1.67$\times 10^{-6}$&-0.02&0.19&-0.02&9.73$\times 10^{-7}$&0.63&0.06&-0.02\\
   		\end{tabular}
   	\end{ruledtabular}
   	\label{fitparameters1}
   \end{table}
   \begin{table}
   	\centering
   	\caption{Fit function parameters for the form factors of the $\Omega_{b}\rightarrow \Omega_{c}^{*}\ell\bar{\nu}_{\ell}$ semileptonic weak transition, considering uncertainties from both auxiliary and input QCD parameters.}
   	\begin{ruledtabular}
   		\begin{tabular}{|c|c|c|c|c|c|c|c|c|}
   			&$F_{1}(q^{2})$&$F_{2}(q^{2})$&$F_{3}(q^{2})$&$F_{4}(q^{2})$&$G_{1}(q^{2})$&$G_{2}(q^{2})$&$G_{3}(q^{2})$&$G_{4}(q^{2})$\\
   			\hline
   			$\mathcal{F}(q^{2}=0)$&$0.043\pm0.001$&$-1.54\pm0.23$&$1.54\pm0.40$&$-0.77\pm0.28$&$-0.042\pm0.001$&$-0.59\pm0.25$&$-0.69\pm0.25$&$0.77\pm0.28$\\
   			$a_{1}$&0.097&1.45&2.47&2.38&0.09&1.10&4.54&2.38\\
   			$a_{2}$&0.01&0.26&2.24&1.74&0.01&0.02&12.25&1.74\\
   			$a_{3}$&-0.001&0.11&-1.10&-0.29&-0.001&0.02&-22.01&-0.29\\
   			$a_{4}$&0.00008&0.03&0.36&-0.08&0.00009&0.03&18.65&-0.08\\
   		\end{tabular}
   	\end{ruledtabular}
   	\label{fitparameters2}
   \end{table}
    In QCD sum rules method, determination of the form factors depends on the Lorentz structures which are selected in a way that result in the least possible uncertainties for the form factors through the working areas of the Borel parameters, continuum thresholds, and mixing parameter $x$. Hence, in this study, for the $\Omega_{b}^{*}\rightarrow \Omega_{c}\ell\bar{\nu}_{\ell}$ transition, we utilize the structures $g_{\mu\nu}\slashed{p}^{\prime}\gamma_5, p^{\prime}_{\nu}\gamma_{\mu}\slashed{p}\slashed{p}^{\prime}\gamma_{5}, p_{\mu}p^{\prime}_{\nu}\slashed{p}^{\prime}\gamma_{5}, p^{\prime}_{\mu}p^{\prime}_{\nu}\slashed{p}\slashed{p}^{\prime}\gamma_{5}, g_{\mu\nu}\slashed{p}, p^{\prime}_{\nu}\gamma_{\mu}\slashed{p}^{\prime}, p_{\mu}p^{\prime}_{\nu}\slashed{p},$ and $p^{\prime}_{\mu}p^{\prime}_{\nu}\slashed{p}\slashed{p}^{\prime}$ which indicate more stability and less dependency on variations of the auxiliary parameters. The structures, $g_{\mu\nu}\slashed{p}^{\prime}\gamma_5, p^{\prime}_{\nu}\gamma_{\mu}\slashed{p}\slashed{p}^{\prime}\gamma_{5}, p_{\mu}p^{\prime}_{\nu}\slashed{p}^{\prime}\gamma_{5}, p^{\prime}_{\mu}p^{\prime}_{\nu}\slashed{p}\slashed{p}^{\prime}\gamma_{5}, g_{\mu\nu}\slashed{p}, p^{\prime}_{\nu}\gamma_{\mu}\slashed{p}^{\prime}, p_{\mu}p^{\prime}_{\nu}\slashed{p},$ and $p^{\prime}_{\mu}p^{\prime}_{\nu}\slashed{p}\slashed{p}^{\prime}$, are related to the eight form factors $F_{1}, F_{2}, F_{3}, F_{4}, G_{1}, G_{2}, G_{3}$, and $G_{4}$, respectively, that are presented in Table \ref{fitparameters1}. The uncertainties of the form factors at $q^2=0$ are based on the uncertainties emerge from the determination of the practical domains for the helping parameters and errors of the other input QCD and physical parameters which are listed in Table \ref{inputparameters}.
    Regarding these uncertainties, the illustrations of the behavior of these form factors in respect of $q^2$ at the central values of the auxiliary parameters, $M^2, M'^2, s_0, s'_0$, and the Ioffe point, $x=-0.71$, as an example, are presented in Fig. \ref{ffq2error}. The increasing form factors, by rising $q^2$, depict the consistency with the weak transition's expectation. We proceed by utilizing these form factors and fit functions to determine the decay rates in all leptonic channels.
    \section{computation of semileptonic and nonleptonic weak decay width}\label{decaywidth} 
    \subsection{$\Omega^{*}_{b}\rightarrow \Omega_{c}$ transition}\label{decayA}
    Having obtained the fit functions of the responsible form factors for the semileptonic  $\Omega^{*}_{b}\rightarrow \Omega_{c}\ell\bar{\nu}_{\ell}$ transition, we employ the helicity amplitudes, in both vector and axial vector representations, $H^{V,A}_{\lambda_{2},\lambda_{W}}$, to compute the decay rates of these transitions in three leptonic modes. Vector and axial vector helicity amplitudes are derived in terms of the vector, $F_{i}^{V}\equiv F_{i} (i=1,2,3,4)$, and the axial vector, $F_{i}^{A}\equiv G_{i}$, form factors, respectively, as presented in Eqs. \eqref{vector} and \eqref{axialvector}, where $\lambda_{W}=t, 0, \pm 1$ is the helicity of the off shell W boson and $\lambda_{2}=\pm\frac{1}{2}$ relates to the final baryon's helicity \cite{Faessler:2009xn,Gutsche:2018nks}. Utilization of $H^{V,A}_{\lambda_{2},\lambda_{W}}=M_{\mu}^{V,A}\bar{\epsilon}^{*\mu}(\lambda_{W})$ enables us to derive the vector and axial vector components of the helicity amplitude where $M_{\mu}^{V,A}$, as stated in Eq. \eqref{formfactor}, represents the vector and axial vector transition currents, $J_{\mu}^{V,A}$, placed between the initial and final hadronic modes and $\bar{\epsilon}^{*\mu}$ indicates the polarization vector of W boson. Therefore, the helicity amplitude components are written as,
     \begin{eqnarray}\label{vector}
    	&H^{V}_{\frac{1}{2},t}=-\sqrt{\frac{2}{3}}\alpha^{V}_{\frac{1}{2},t}(\omega-1)\Big(F_{1}^{V}M_{2}+F_{2}^{V}M_{+}-F_{3}^{V}\frac{M_{1}}{M_{2}}(M_{1}-M_{2}\omega)-F_{4}^{V}\frac{q^{2}}{M_{2}}\Big),\nonumber\\
    	&H^{V}_{\frac{1}{2},0}=-\sqrt{\frac{2}{3}}\alpha^{V}_{\frac{1}{2},0}\Big(F_{1}^{V}(M_{1}-M_{2}\omega)+F_{2}^{V}(\omega+1)M_{-}-F_{3}^{V}(\omega^{2}-1)M_{1}\Big),\nonumber\\
    	&H^{V}_{\frac{1}{2},1}=\frac{1}{\sqrt{6}}\alpha^{V}_{\frac{1}{2},1}\Big(\,F_{1}^{V}+2F_{2}^{V}(\omega+1)\Big),\nonumber\\
    	&H^{V}_{\frac{1}{2},-1}=\frac{1}{\sqrt{2}}\alpha^{V}_{\frac{1}{2},1}F_{1}^{V},
    \end{eqnarray} 
    and
    	\begin{eqnarray}\label{axialvector}
    	&H^{A}_{\frac{1}{2},t}=\sqrt{\frac{2}{3}}\alpha^{A}_{\frac{1}{2},t}(\omega+1)\Big(F_{1}^{A}M_{2}-F_{2}^{A}M_{-}-F_{3}^{A}\frac{M_{1}}{M_{2}}(M_{1}-M_{2}\omega)-F_{4}^{A}\frac{q^{2}}{M_{2}}\Big),\nonumber\\
    	&H^{A}_{\frac{1}{2},0}=-\sqrt{\frac{2}{3}}\alpha^{A}_{\frac{1}{2},0}\Big(\,-F_{1}^{A}(M_{1}-M_{2}\omega)+F_{2}^{A}(\omega-1)M_{+}+F_{3}^{A}(\omega^{2}-1)M_{1}\Big),\nonumber\\
    	&H^{A}_{\frac{1}{2},1}=\frac{1}{\sqrt{6}}\alpha^{A}_{\frac{1}{2},1}\Big(-F_{1}^{A}+2F_{2}^{A}(\omega-1)\Big),\nonumber\\
    	&H^{A}_{\frac{1}{2},-1}=-\frac{1}{\sqrt{2}}\alpha^{A}_{\frac{1}{2},1}F_{1}^{A},
    \end{eqnarray} 
    by employing the following definitions,
     \begin{eqnarray}\label{definition}
    	&&\alpha^{V}_{\frac{1}{2},t}=\alpha^{A}_{\frac{1}{2},0}=\sqrt{\frac{2M_{1}M_{2}(\omega+1)}{q^{2}}},\nonumber\\
    	&&\alpha^{V}_{\frac{1}{2},0}=\alpha^{A}_{\frac{1}{2},t}=\sqrt{\frac{2M_{1}M_{2}(\omega-1)}{q^{2}}},\nonumber\\
    	&&\alpha^{V}_{\frac{1}{2},1}=2\sqrt{M_{1}M_{2}(\omega-1)},\nonumber\\
    	&&\alpha^{A}_{\frac{1}{2},1}=2\sqrt{M_{1}M_{2}(\omega+1)},
    \end{eqnarray}
    and 
    \[ M_{\pm}=M_{1}\pm M_{2}, ~~~\omega=\frac{M_{1}^{2}+M_{2}^{2}-q^{2}}{2M_{1}M_{2}},
    \] 
    where $M_1$ and $M_2$ denote the masses of the initial and final baryons, respectively. In addition, we have 
    \begin{equation}
    	H^{V}_{-\lambda_{2},-\lambda_{W}}=-	H^{V}_{\lambda_{2},\lambda_{W}}, ~~~H^{A}_{-\lambda_{2},-\lambda_{W}}=H^{A}_{\lambda_{2},\lambda_{W}},
    \end{equation} 
    for the negative helicity amplitudes. Subsequently, regarding the vector-axial vector structure of the weak transitions, the helicity amplitude is obtained as $H_{\lambda_{2},\lambda_{W}}=H^{V}_{\lambda_{2},\lambda_{W}}-\,H^{A}_{\lambda_{2},\lambda_{W}}$. Ultimate results of the helicity amplitudes provide us with the determination of decay widths for the semileptonic $\Omega^{*}_{b}\rightarrow \Omega_{c}\ell\bar{\nu}_{\ell}$ transitions in all lepton channels, using the following formula \cite{Faessler:2009xn},
     \begin{equation}
    	\Gamma_{\Omega_{b}^{*}\rightarrow \Omega_{c}\ell \bar{\nu}_{\ell}}=\frac{1}{2}\frac{G_{F}^{2}|V_{cb}|^{2}}{192\pi^{3}}\frac{M_{\Omega_{c}}}{M_{\Omega_{b}^{*}}^{2}}\int_{m_{\ell}^{2}}^{(M_{\Omega_{b}^{*}}-M_{\Omega_{c}})^{2}}\frac{\mathrm{d}q^{2}}{q^{2}}(q^{2}-m_{\ell}^{2})^{2}\sqrt{\omega^{2}-1}~\mathcal{H}_{\frac{3}{2}\rightarrow \frac{1}{2}},
    \end{equation}
    in which $m_{\ell}$ denotes the lepton mass and a combination of helicity amplitudes is incorporated in ${\cal H}_{\frac{3}{2}\rightarrow\frac{1}{2}}$,
     \begin{eqnarray}
    	&&  \mathcal{H}_{\frac{3}{2} \rightarrow \frac{1}{2}}=| H_{\frac{1}{2},1}|^{2}+| H_{-\frac{1}{2},-1}|^{2}+| H_{\frac{1}{2},-1}|^{2}+| H_{-\frac{1}{2},1}|^{2}+| H_{\frac{1}{2},0}|^{2}+| H_{-\frac{1}{2},0}|^{2}+\nonumber\\
    	&&\frac{m_{\ell}^{2}}{2q^{2}}\bigg(3| H_{\frac{1}{2},t}|^{2}+3| H_{-\frac{1}{2},t}|^{2}+| H_{\frac{1}{2},1}|^{2}+| H_{-\frac{1}{2},-1}|^{2}+| H_{\frac{1}{2},-1}|^{2}+| H_{-\frac{1}{2},1}|^{2}+| H_{\frac{1}{2},0}|^{2}+| H_{-\frac{1}{2},0}|^{2}\bigg).
    \end{eqnarray}

    The obtained results of the decay widths are summarized in Table \ref{decaywidth20}. 
     \begin{table}[h]
    	\centering
    	\caption{Decay widths of the semileptonic  $\Omega_{b}^{*}\rightarrow \Omega_{c}\ell\bar{\nu}_{\ell}$ transitions at different leptonic channels in units of GeV, considering uncertainties from both auxiliary and input QCD parameters.}
    	\begin{ruledtabular}
    		\begin{tabular}{|c|c|c|}
    			& & \\
    			$\Gamma[\Omega_{b}^{*}\rightarrow \Omega_{c}e\bar{\nu}_{e}]\times 10^{14}$&$\Gamma[\Omega_{b}^{*}\rightarrow \Omega_{c}\mu\bar{\nu}_{\mu}]\times 10^{14}~~~~~~~~$&$\Gamma[\Omega_{b}^{*}\rightarrow \Omega_{c}\tau\bar{\nu}_{\tau}]\times 10^{15}~~~~~~~~$\\
    			\hline
    			& & \\
    			$1.54^{+0.41}_{-0.36}$&$1.53^{+0.39}_{-0.36}$&$3.15^{+0.79}_{-0.66}$\\
    			& & \\
    		\end{tabular}
    	\end{ruledtabular}
    	\label{decaywidth20}
    \end{table} 
    Our findings indicate a satisfying consistency with the kinematical requirements for which the decay widths are supposed to decrease by rising lepton mass. In order to eliminate some systematic uncertainties, the ratio of decay widths in $\tau$ to $e/\mu$ channel is presented as,
     \begin{equation}\label{Ratio}
    	R_{\Omega_{b}^{*}}=\frac{\Gamma[\Omega_{b}^{*}\rightarrow \Omega_{c}\tau\bar{\nu}_{\tau}]}{\Gamma[\Omega_{b}^{*}\rightarrow \Omega_{c}e(\mu)\bar{\nu}_{e(\mu)}]}=0.21\pm0.02 .
    \end{equation}
    These findings may guide subsequent experiments in exploring these decay channels and probing the hadronic structures of b-heavy baryons. A detailed investigation of b-heavy baryons enables us to evaluate the SM predictions and search for signals of new physics.

    In addition to the semileptonic weak decays, the nonleptonic decays of heavy baryons receive considerable attention since a detailed analysis of these decays provide us with valuable information on heavy baryons' properties and the main features and parameters of QCD. In the following, we study the two-body nonleptonic weak decays of the b-heavy baryon, $\Omega_{b}^{*}$, focusing on W-emission contribution which is known as tree graph contribution.\\Indeed, we investigate the two-body nonleptonic transitions of $\Omega_{b}^{*}$ to  $\Omega_{c}$ with emitting a pseudoscalar meson ($\pi^{-}, K^{-}, D^{-},$ or  $D_{s}^{-}$) or a vector meson ($\rho^{-}, K^{*-}, D^{*-},$ or  $D^{*-}_{s}$). 
    
    Naive factorization approach is the popular method to deal with the nonleptonic weak decays approximately. In this approach, the matrix elements of  four-quark operators are computed by factorizing them into a product of two independent currents. However, in baryonic sector of the nonleptonic weak decays, because of the complicated structure of baryons, in contrast to mesons, the strong interaction of diquarks, final-state interactions, and strong phases have considerable importance and the factorization approximation becomes partially justified. Moreover, the nonfactorizable contributions including W-exchange ones can be as significant as the factorizable terms. Nevertheless, bottom baryons,are particularly different. In these cases, the factorizable W-emission contributions dominate over the nonfactorizable W-exchange terms. Also, a number of bottom baryon decays obtains contributions only from factorizable terms. Consequently, factorization approximation, as a rough estimate, can be employed to describe the nonleptonic weak decays of bottom baryons.
    
    Under the approximate factorization, the hadronic transition matrix element is given as \cite{Li:2021kfb, Chua:2018lfa,Meiser:2024zea,Liu:2023dvg,Wang:2023uea,Wang:2024ozz,Sharma:2009zzg,Jia:2024pyb,Ivanov:2021huf,Zhao:2022vfr},
    \begin{equation}
    \mathcal{A}(\Omega^{*}_{b}\rightarrow \Omega_{c} M)=\frac{G_{F}}{\sqrt{2}}V_{cb}V_{qq'}^{*}C_{eff}\langle\Omega_{c}|\bar{c}\gamma_{\mu}(1-\gamma_{5})b|\Omega^{*}_{b}\rangle~\langle M|\bar{q'}\gamma_{\mu}(1-\gamma_{5})q|0\rangle,  
    \end{equation} 
     where $V_{cb}$ and  $V_{qq'}$ are the CKM matrix elements and $C_{eff}$ denotes the effective scale-dependent Wilson coefficient which is presented as $C_{1}(\mu)+\frac{C_2(\mu)}{N_c}$ that $N_c$ is the number of colors \cite{Gutsche:2018utw,Chua:2018lfa,Zhao:2022vfr}. The matrix element $\langle\Omega_{c}|\bar{c}\gamma_{\mu}(1-\gamma_{5})b|\Omega^{*}_{b}\rangle$ is given by Eq. \eqref{formfactor}, in terms of form factors, and the meson transition term is parametrized as,
     \begin{eqnarray}
     &&\langle P|\bar{q'}\gamma_{\mu}(1-\gamma_{5})q|0\rangle=iq^{P}_{\mu}f_{P},\notag\\
     &&\langle V|\bar{q'}\gamma_{\mu}(1-\gamma_{5})q|0\rangle=m_V f_{V}\epsilon^{*}_{\mu},
     \end{eqnarray}
     where P and V stand for the pseudoscalar and vector mesons. $f_P$ and $q_{\mu}^{P}$ represent the decay constant and momentum of the pseudoscalar meson, respectively, while $f_V$ indicates the vector meson decay constant and  $\epsilon^{*}_{\mu}$ is the polarization vector. Subsequently, we identify the vector and axial vector helicity amplitudes, $H_{\lambda_2\lambda_M}^{V/A}$, in terms of the form factors, $F_{i}^{V/A}$, where $\lambda_2=\pm\frac{1}{2}$ and $\lambda_M=t,0,\pm1$ are the helicity components of the final baryon and the meson, respectively. Before proceeding with the computation of the following expression,
     \begin{equation}
     	H_{\lambda_2\lambda_M}=\langle\Omega_{c}(p',\lambda_2)|\bar{c}\gamma_{\mu}(1-\gamma_{5})b|\Omega^{*}_{b}(p,\lambda_1)\rangle\epsilon^{*\mu}(\lambda_M)=H_{\lambda_2\lambda_M}^{V}-H_{\lambda_2\lambda_M}^{A},
     \end{equation}
      it is essential to determine the transition form factors by employing the on-shell condition $q^2=m_{Meson}^2$ where $q$ is the transferred momentum and $m$ is the mass of the meson. Hence, we present the obtained results of the form factors for the nonleptonic  $\Omega^{*}_{b}\rightarrow \Omega_{c} M$ weak decays, as  $\frac{3}{2}\rightarrow\frac{1}{2}$ transitions, in Table \ref{formfactornon32}.
      \begin{table}
      	\centering
      	\caption{Transition form factors of the nonleptonic $\Omega_{b}^{*}\rightarrow \Omega_{c}M$ weak decays with emitting a pseudoscalar or vector meson at $q^2=m_{Meson}^{2}$, considering uncertainties from both auxiliary and input QCD parameters.}
      	\begin{ruledtabular}
      		\begin{tabular}{ccccccccc}
      			transition&$F_{1}(q^{2})$&$F_{2}(q^{2})$&$F_{3}(q^{2})$&$F_{4}(q^{2})$&$G_{1}(q^{2})$&$G_{2}(q^{2})$&$G_{3}(q^{2})$&$G_{4}(q^{2})$\\
      			\hline
      			$\Omega^{*}_{b}\rightarrow \Omega_{c}\pi^{-}$&$-0.08\pm0.01$&$-1.26\pm0.28$&$-0.39\pm0.10$&$-0.31\pm0.10$&$-0.06\pm0.01$&$0.39\pm0.12$&$-0.41\pm0.09$&$-0.31\pm0.09$\\
      			$\Omega^{*}_{b}\rightarrow \Omega_{c}K^{-}$&$-0.08\pm0.01$&$-1.27\pm0.27$&$-0.39\pm0.12$&$-0.32\pm0.12$&$-0.06\pm0.01$&$0.40\pm0.13$&$-0.42\pm0.10$&$-0.32\pm0.07$\\
      			$\Omega^{*}_{b}\rightarrow \Omega_{c}D^{-}$&$-0.09\pm0.03$&$-1.50\pm0.39$&$-0.45\pm0.12$&$-0.40\pm0.09$&$-0.06\pm0.01$&$0.50\pm0.10$&$-0.48\pm0.06$&$-0.40\pm0.09$\\
      			$\Omega^{*}_{b}\rightarrow \Omega_{c}D_{s}^{-}$&$-0.09\pm0.02$&$-1.53\pm0.34$&$-0.46\pm0.09$&$-0.42\pm0.10$&$-0.06\pm0.01$&$0.51\pm0.13$&$-0.48\pm0.13$&$-0.42\pm0.09$\\
      			$\Omega^{*}_{b}\rightarrow \Omega_{c}\rho^{-}$&$-0.08\pm0.02$&$-1.29\pm0.31$&$-0.40\pm0.07$&$-0.32\pm0.09$&$-0.06\pm0.01$&$0.41\pm0.12$&$-0.42\pm0.09$&$-0.32\pm0.07$\\
      			$\Omega^{*}_{b}\rightarrow \Omega_{c}K^{*-}$&$-0.08\pm0.02$&$-1.30\pm0.27$&$-0.40\pm0.10$&$-0.33\pm0.09$&$-0.06\pm0.01$&$0.41\pm0.07$&$-0.43\pm0.09$&$-0.33\pm0.06$\\
      			$\Omega^{*}_{b}\rightarrow \Omega_{c}D^{*-}$&$-0.09\pm0.02$&$-1.54\pm0.34$&$-0.46\pm0.10$&$-0.42\pm0.09$&$-0.06\pm0.01$&$0.52\pm0.15$&$-0.49\pm0.07$&$-0.42\pm0.06$\\
      			$\Omega^{*}_{b}\rightarrow \Omega_{c}D_{s}^{*-}$&$-0.09\pm0.03$&$-1.57\pm0.37$&$-0.47\pm0.09$&$-0.44\pm0.09$&$-0.06\pm0.01$&$0.53\pm0.15$&$-0.49\pm0.09$&$-0.44\pm0.09$\\
      		\end{tabular}
      	\end{ruledtabular}
      	\label{formfactornon32}
      \end{table}

  The vector and axial vector  helicity amplitudes are derived in terms of these form factors as follows,
  \begin{eqnarray}\label{helicityamplitudenon}
  H_{\frac{1}{2},0}^{V,A}&=&\pm i\sqrt{\frac{2}{3}}\frac{\sqrt{2M_1M_2(\omega\mp1)}}{\hat{q}}\Big[(\mp1+r)(\omega\pm1)F_2^{V,A}+r(\frac{M_1}{M_2})(\omega^2-1)F_3^{V,A}\nonumber\\&+&(\omega^2-1)(r\frac{M_1}{M_2}-1)F_4^{V,A}+(r\omega-1)F_1^{V,A}\Big],\nonumber\\	
  H_{\frac{1}{2},t}^{V,A}&=&\mp i\sqrt{\frac{2}{3}}(\omega\mp1)\frac{\sqrt{2M_1M_2(\omega\pm1)}}{\hat{q}}\Big[(\pm1+r)F_2^{V,A}+(r\omega-1)(\frac{M_1}{M_2})F_3^{V,A}\nonumber\\&+&[(r\omega-1)(\frac{M_1}{M_2})-(r-\omega)]F_4^{V,A}+r F_1^{V,A}\Big],\nonumber\\	 
  H_{\frac{1}{2},1}^{V,A}&=&\mp i\sqrt{\frac{1}{3}}\sqrt{2M_1M_2(\omega\mp1)}\Big[2(\pm\omega+1)F_{2}^{V,A}+F_{1}^{V,A}\Big],\nonumber\\
  H_{\frac{1}{2},-1}^{V,A}&=&\mp i\sqrt{2M_1M_2(\omega\mp1)}F_{1}^{V,A},
  \end{eqnarray} 
  where the upper (lower) sign relates to V(A), $r\equiv\frac{M_2}{M_1}$, and $\hat{q}\equiv\frac{\sqrt{q^2}}{M_1}$. The other helicity amplitudes are given by $	H^{V,A}_{-\lambda_{2},-\lambda_{M}}=\mp H^{V,A}_{\lambda_{2},\lambda_{M}}$. 
   Ultimately, the decay widths of the two-body nonleptonic weak transitions for pseudoscalar $(P)$ and vector $(V)$ meson case are expressed, respectively, as follows,
   \begin{eqnarray}
  	\Gamma(\frac{3}{2}\rightarrow\frac{1}{2})=\Gamma(\Omega^{*}_{b}\rightarrow\Omega_{c}P)&=&(\frac{G_F}{\sqrt{2}})^{2}\frac{|\vec{P'}|}{32\pi M_{\Omega_{b}^{*}}^{2}}|V_{cb}V^*_{qq'}|^2C_{eff}^{2}f_{P}^2m_{Meson}^2\bigg(| H_{\frac{1}{2},t}|^{2}+| H_{-\frac{1}{2},t}|^{2}\bigg),\nonumber\\
 	\Gamma(\frac{3}{2}\rightarrow\frac{1}{2})=\Gamma(\Omega^{*}_{b}\rightarrow\Omega_{c}V)&=&(\frac{G_F}{\sqrt{2}})^{2}\frac{|\vec{P'}|}{32\pi M_{\Omega_{b}^{*}}^{2}}|V_{cb}V^*_{qq'}|^2C_{eff}^{2}f_{V}^2m_{Meson}^2\bigg(| H_{\frac{1}{2},1}|^{2}+| H_{-\frac{1}{2},-1}|^{2}\nonumber\\&&+| H_{\frac{1}{2},-1}|^{2}+| H_{-\frac{1}{2},1}|^{2}+| H_{\frac{1}{2},0}|^{2}+| H_{-\frac{1}{2},0}|^{2}\bigg),
  	\end{eqnarray} 
  	where $|\vec{P'}|=\frac{\lambda^{\frac{1}{2}}(M_{\Omega^{*}_{b}}^2,M_{\Omega_{c}}^2, q^2)}{2M_{\Omega^*_b}}$ and $\lambda(M_{\Omega^{*}_{b}}^2,M_{\Omega_{c}}^2,q^2)=M_{\Omega^{*}_{b}}^4+M_{\Omega_{c}}^4+q^4-2M_{\Omega^{*}_{b}}^2M_{\Omega_{c}}^2-2M_{\Omega^{*}_{b}}^2q^2-2M_{\Omega_{c}}^2q^2$.
  	
  	Having obtained the form factors from QCD sum rules method and the helicity amplitudes, decay widths of the considered nonleptonic $\Omega^{*}_{b}\rightarrow\Omega_{c}M$ weak decays are determined as presented in Table \ref{decaywidth32}. Regarding the significant progress, achieved in experiments, these theoretical results, obtained for the first time especially within the QCD sum rules method, can be verified by experiment in future.
  	 \begin{table}
  		\centering
  		\caption{Decay widths of the nonleptonic $\Omega^{*}_{b}\rightarrow\Omega_{c}M$ decays with $M$ indicating a pseudoscalar or vector meson in units of GeV, considering uncertainties from both auxiliary and input QCD parameters.}
  		\begin{ruledtabular}
  			\begin{tabular}{cccc}
  				transition&$\Gamma$(GeV)&transition&$\Gamma$(GeV)\\
  				\hline
  				$\Omega^{*}_{b}\rightarrow \Omega_{c}\pi^{-}$&$(1.26^{+0.34}_{-0.33})\times10^{-15}$&$\Omega^{*}_{b}\rightarrow \Omega_{c}\rho^{-}$&$(4.19^{+1.20}_{-1.14})\times10^{-15}$\\
  				$\Omega^{*}_{b}\rightarrow \Omega_{c}K^{-}$&$(1.09^{+0.28}_{-0.27})\times10^{-16}$&$\Omega^{*}_{b}\rightarrow \Omega_{c}K^{*-}$&$(2.26^{+0.59}_{-0.56})\times10^{-16}$\\
  				$\Omega^{*}_{b}\rightarrow \Omega_{c}D^{-}$&$(6.81^{+1.82}_{-1.73})\times10^{-17}$&$\Omega^{*}_{b}\rightarrow \Omega_{c}D^{*-}$&$(2.67^{+0.65}_{-0.62})\times10^{-16}$\\
  				$\Omega^{*}_{b}\rightarrow \Omega_{c}D_{s}^{-}$&$(1.67^{+0.44}_{-0.42})\times10^{-15}$&$\Omega^{*}_{b}\rightarrow \Omega_{c}D_{s}^{*-}$&$(4.80^{+1.22}_{-1.15})\times10^{-15}$
  			\end{tabular}
  		\end{ruledtabular}
  		\label{decaywidth32}
  	\end{table}
  	In our computations, the CKM matrix elements are taken from the particle data group (PDG) review \cite{ParticleDataGroup:2024cfk},
  	\begin{eqnarray}
  	&&V_{ud}=0.974, V_{cb}=0.040, V_{us}=0.226,\notag\\
  	&&V_{cd}=0.226, V_{cs}=0.973,\notag	
  	\end{eqnarray}
  	and the decay constants of the pseudoscalar and vector mesons are considered as:\\
  		$f_{\pi}=130.2~\mathrm{MeV}$ \cite{Narison:2012xy,ParticleDataGroup:2024cfk},  $f_{K}=155.6~\mathrm{MeV}$\cite{Narison:2012xy,ParticleDataGroup:2024cfk},  $f_{D}=203.7~\mathrm{MeV} $ \cite{Narison:2012xy},  $f_{D_s}=246.0~\mathrm{MeV}$  \cite{Narison:2012xy},
  	 $f_{\rho}=218.3~\mathrm{MeV}$ \cite{Gutsche:2018utw},  $f_{K^*}=210~\mathrm{MeV}$ \cite{Li:2021kfb},  $f_{D^*}=245\pm20~\mathrm{MeV}$ \cite{Becirevic:1998ua},  $f_{D_s^*}=(272\pm16)~\mathrm{MeV}$ \cite{Becirevic:1998ua}.\\
  	The effective Wilson coefficient, for the bottom quark decay, at the scale of $\mu_b=4.2~\mathrm{GeV}$ is $C_{eff}=1.02$ and the masses of the mesons are taken from PDG \cite{ParticleDataGroup:2024cfk}. The next section is devoted to the other transition, $\Omega_{b}\rightarrow\Omega^{*}_{c}$, and the semileptonic and nonleptonic weak decays of this transition, in different channels, are investigated.  
  	\subsection{$\Omega_{b}\rightarrow\Omega^{*}_{c}$ transition}\label{decayB}
  	In Sec. \ref{numerical}, we employed the QCD ‌sum rules method to obtain the form factors of the semileptonic $\Omega_{b}\rightarrow\Omega^{*}_{c}\ell\bar{\nu}_{\ell}$ transition. These form factors, as the main features of the method, presented in Table \ref{fitparameters2}, are utilized to compute the observable quantities such as decay widths and branching fractions in different decay modes. In a similar way to the previous section, it is required to derive the helicity amplitudes, associated with the $\frac{1}{2}\rightarrow\frac{3}{2}$ process, in terms of these transition form factors as follows, 
     \begin{eqnarray}
    	&&H^{V}_{\frac{1}{2},t}=-\sqrt{\frac{2}{3}}\alpha^{V}_{\frac{1}{2},t}(\omega-1)\Big(F_{1}^{V}M_{1}-F_{2}^{V}M_{+}+F_{3}^{V}\frac{M_{2}}{M_{1}}(M_{1}\omega-M_{2})+F_{4}^{V}\frac{q^{2}}{M_{1}}\Big),\nonumber\\
    	&&H^{V}_{\frac{1}{2},0}=-\sqrt{\frac{2}{3}}\alpha^{V}_{\frac{1}{2},0}\Big(F_{1}^{V}(M_{1}\omega-M_{2})-F_{2}^{V}(\omega+1)M_{-}+F_{3}^{V}(\omega^{2}-1)M_{2}\Big),\nonumber\\
    	&&H^{V}_{\frac{1}{2},1}=\frac{1}{\sqrt{6}}\alpha^{V}_{\frac{1}{2},1}\Big(F_{1}^{V}-2F_{2}^{V}(\omega+1)\Big),\nonumber\\
    	&&H^{V}_{\frac{3}{2},1}=-\frac{1}{\sqrt{2}}\alpha^{V}_{\frac{1}{2},1}F_{1}^{V},
    \end{eqnarray}
    and
    	\begin{eqnarray}
    	&&H^{A}_{\frac{1}{2},t}=\sqrt{\frac{2}{3}}\alpha^{A}_{\frac{1}{2},t}(\omega+1)\Big(F_{1}^{A}M_{1}+F_{2}^{A}M_{-}+F_{3}^{A}\frac{M_{2}}{M_{1}}(M_{1}\omega-M_{2})+F_{4}^{A}\frac{q^{2}}{M_{1}}\Big),\nonumber\\
    	&&H^{A}_{\frac{1}{2},0}=\sqrt{\frac{2}{3}}\alpha^{A}_{\frac{1}{2},0}\Big(F_{1}^{A}(M_{1}\omega-M_{2})+F_{2}^{A}(\omega-1)M_{+}+F_{3}^{A}(\omega^{2}-1)M_{2}\Big),\nonumber\\
    	&&H^{A}_{\frac{1}{2},1}=\frac{1}{\sqrt{6}}\alpha^{A}_{\frac{1}{2},1}\Big(F_{1}^{A}-2F_{2}^{A}(\omega-1)\Big),\nonumber\\
    	&&H^{A}_{\frac{3}{2},1}=\frac{1}{\sqrt{2}}\alpha^{A}_{\frac{1}{2},1}F_{1}^{A},
    \end{eqnarray} 
   where the same definitions as Def. \eqref{definition} are considered. Also the negative helicity amplitudes are determined by $H^{V}_{-\lambda_{2},-\lambda_{W}}=-	H^{V}_{\lambda_{2},\lambda_{W}}$ and $H^{A}_{-\lambda_{2},-\lambda_{W}}=H^{A}_{\lambda_{2},\lambda_{W}}$. Ultimately, the obtained results of the form factors and the helicity amplitudes enable us to compute the decay rates of the semileptonic $\Omega_{b}\rightarrow\Omega^{*}_{c}\ell\bar{\nu}_{\ell}$ weak transitions in three lepton channels, employing the formula \cite{Faessler:2009xn}, 
     \begin{equation}
    	\Gamma_{\Omega_{b}\rightarrow \Omega_{c}^{*}\ell \bar{\nu}_{\ell}}=\frac{G_{F}^{2}|V_{cb}|^{2}}{192\pi^{3}}\frac{M_{\Omega^{*}_{c}}}{M_{\Omega_{b}}^{2}}\int_{m_{\ell}^{2}}^{(M_{\Omega_{b}}-M_{\Omega^{*}_{c}})^{2}}\frac{\mathrm{d}q^{2}}{q^{2}}(q^{2}-m_{\ell}^{2})^{2}\sqrt{\omega^{2}-1}~\mathcal{H}_{\frac{1}{2}\rightarrow \frac{3}{2}},
    \end{equation}
    in which $\mathcal{H}_{\frac{1}{2}\rightarrow \frac{3}{2}}$ includes,
     \begin{eqnarray}
    	&&  \mathcal{H}_{\frac{1}{2} \rightarrow \frac{3}{2}}=| H_{\frac{1}{2},1}|^{2}+| H_{-\frac{1}{2},-1}|^{2}+| H_{\frac{3}{2},1}|^{2}+| H_{-\frac{3}{2},-1}|^{2}+| H_{\frac{1}{2},0}|^{2}+| H_{-\frac{1}{2},0}|^{2}+\nonumber\\
    	&&\frac{m_{\ell}^{2}}{2q^{2}}\bigg(3| H_{\frac{1}{2},t}|^{2}+3| H_{-\frac{1}{2},t}|^{2}+| H_{\frac{1}{2},1}|^{2}+| H_{-\frac{1}{2},-1}|^{2}+| H_{\frac{3}{2},1}|^{2}+| H_{-\frac{3}{2},-1}|^{2}+| H_{\frac{1}{2},0}|^{2}+| H_{-\frac{1}{2},0}|^{2}\bigg).
    \end{eqnarray}
    The ultimate results of the semileptonic decay widths and branching ratios in three lepton channels are presented in Table \ref{decaywidth22}. The lifetime of the $\Omega_b$ baryon is $\tau_{\Omega_b}=(1.64\pm0.16)\times 10^{-12}$s which is taken from PDG  \cite{ParticleDataGroup:2024cfk}. 
     \begin{table}[h!]
    	\centering
    	\caption{Decay widths and branching ratios of the semileptonic  $\Omega_{b}\rightarrow \Omega^{*}_{c}(2766)\ell\bar{\nu}_{\ell}$ transitions at different lepton channels, considering uncertainties from both auxiliary and input QCD parameters.}
    	\begin{ruledtabular}
    		\begin{tabular}{cccc}
    			transition&$\Omega_{b}\rightarrow \Omega^{*}_{c}e\bar{\nu}_{e}$&	$\Omega_{b}\rightarrow \Omega^{*}_{c}\mu\bar{\nu}_{\mu}$&$\Omega_{b}\rightarrow \Omega^{*}_{c}\tau\bar{\nu}_{\tau}$\\
    			\hline
    			& & &\\
    			$\Gamma(\mathrm{GeV})$&$(4.30^{+1.67}_{-1.72})\times 10^{-14}$&$(4.27^{+1.65}_{-1.72})\times 10^{-14}$&$(5.92^{+2.26}_{-2.36})\times 10^{-15}$\\
    			& & &\\
    			$Br$&$(10.72^{+4.24}_{-4.28})\times10^{-2}$&$(10.64^{+4.04}_{-4.27})\times10^{-2}$&$(1.48^{+0.56}_{-0.58})\times10^{-2}$\\
    		\end{tabular}
    	\end{ruledtabular}
    	\label{decaywidth22}
    \end{table}
    The ratio of the branching fractions in $\tau$ to $e/\mu$ mode is determined as follows,
     \begin{equation}\label{Ratio2}
    	R_{\Omega_{b}}=\frac{Br[\Omega_{b}\rightarrow \Omega^{*}_{c}\tau\bar{\nu}_{\tau}]}{Br[\Omega_{b}\rightarrow \Omega^{*}_{c}e(\mu)\bar{\nu}_{e(\mu)}]}=0.14\pm0.02 .
    \end{equation}

    In the nonleptonic decay sector, we investigate the two-body nonleptonic weak transitions of $\Omega_{b}$ to  $\Omega^{*}_{c}$ with emitting a pseudoscalar meson ($\pi^{-}, K^{-}, D^{-},$ or  $D_{s}^{-}$) or vector meson ($\rho^{-}, K^{*-}, D^{*-},$ or  $D^{*-}_{s}$). The decay amplitude, related to the nonleptonic transition, is expressed as, 
     \begin{equation}
    	\mathcal{A}(\Omega_{b}\rightarrow \Omega^{*}_{c} M)=\frac{G_{F}}{\sqrt{2}}V_{cb}V_{qq'}^{*}C_{eff}\langle\Omega^{*}_{c}|\bar{c}\gamma_{\mu}(1-\gamma_{5})b|\Omega_{b}\rangle~\langle M|\bar{q'}\gamma_{\mu}(1-\gamma_{5})q|0\rangle.  
    \end{equation} 
    As previously mentioned, the vector and axial vector helicity amplitudes are provided in terms of the transition form factors. By applying the on-shell condition $q^2=m^2_{Meson}$, the nonleptonic form factors of the $\Omega_{b}\rightarrow \Omega^{*}_{c} M$ weak decays, as $\frac{1}{2}\rightarrow\frac{3}{2}$ transitions, are determined. These results are presented in Table \ref{formfactornon12}.
    \begin{table}
    	\centering
    	\caption{Transition form factors of the nonleptonic $\Omega_{b}\rightarrow \Omega^{*}_{c}M$ weak decays with emitting a pseudoscalar or  vector meson at $q^2=m_{Meson}^{2}$, considering uncertainties from both auxiliary and input QCD parameters.}
    	\begin{ruledtabular}
    		\begin{tabular}{ccccccccc}
    			transition&$F_{1}(q^{2})$&$F_{2}(q^{2})$&$F_{3}(q^{2})$&$F_{4}(q^{2})$&$G_{1}(q^{2})$&$G_{2}(q^{2})$&$G_{3}(q^{2})$&$G_{4}(q^{2})$\\
    			\hline
    			$\Omega_{b}\rightarrow \Omega^{*}_{c}\pi^{-}$&$0.04\pm0.01$&$-1.54\pm0.34$&$1.54\pm0.27$&$-0.77\pm0.16$&$-0.04\pm0.01$&$-0.59\pm0.10$&$-0.69\pm0.13$&$0.77\pm0.09$\\
    			$\Omega_{b}\rightarrow \Omega^{*}_{c}K^{-}$&$0.04\pm0.01$&$-1.56\pm0.31$&$1.57\pm0.33$&$-0.78\pm0.16$&$-0.04\pm0.01$&$-0.60\pm0.12$&$-0.71\pm0.15$&$0.78\pm0.07$\\
    			$\Omega_{b}\rightarrow \Omega^{*}_{c}D^{-}$&$0.04\pm0.01$&$-1.79\pm0.39$&$1.97\pm0.40$&$-0.97\pm0.21$&$-0.04\pm0.01$&$-0.66\pm0.12$&$-1.05\pm0.15$&$0.97\pm0.09$\\
    			$\Omega_{b}\rightarrow \Omega^{*}_{c}D_{s}^{-}$&$0.04\pm0.01$&$-1.82\pm0.40$&$2.03\pm0.39$&$-1.00\pm0.22$&$-0.04\pm0.01$&$-0.67\pm0.09$&$-1.10\pm0.13$&$1.00\pm0.09$\\
    			$\Omega_{b}\rightarrow \Omega^{*}_{c}\rho^{-}$&$0.04\pm0.01$&$-1.58\pm0.31$&$1.61\pm0.37$&$-0.80\pm0.21$&$-0.04\pm0.01$&$-0.60\pm0.10$&$-0.74\pm0.18$&$0.80\pm0.07$\\
    			$\Omega_{b}\rightarrow \Omega^{*}_{c}K^{*-}$&$0.04\pm0.01$&$-1.59\pm0.30$&$1.63\pm0.36$&$-0.81\pm0.18$&$-0.04\pm0.01$&$-0.61\pm0.09$&$-0.76\pm0.15$&$0.81\pm0.06$\\
    			$\Omega_{b}\rightarrow \Omega^{*}_{c}D^{*-}$&$0.04\pm0.01$&$-1.83\pm0.34$&$2.05\pm0.36$&$-1.01\pm0.24$&$-0.04\pm0.01$&$-0.67\pm0.18$&$-1.12\pm0.22$&$1.01\pm0.06$\\
    			$\Omega_{b}\rightarrow \Omega^{*}_{c}D_{s}^{*-}$&$0.04\pm0.01$&$-1.86\pm0.36$&$2.11\pm0.42$&$-1.04\pm0.24$&$-0.04\pm0.01$&$-0.68\pm0.13$&$-1.17\pm0.22$&$1.04\pm0.16$\\
    		\end{tabular}
    	\end{ruledtabular}
    	\label{formfactornon12}
    \end{table}
     The vector and axial vector helicity amplitudes, in terms of the form factors, are expressed as,
     \begin{eqnarray}
     	&H^{V}_{\frac{1}{2},t}=-\sqrt{\frac{2(M^2_{+}-q^2)}{3q^2}}(\frac{M^2_{-}-q^2}{2M_{1}M_{2}})\Big(F_{1}^{V}M_{1}-F_{2}^{V}M_{+}+F_{3}^{V}\frac{M_{+}M_{-}-q^2}{2M_1}+F_{4}^{V}\frac{q^{2}}{M_{1}}\Big),\nonumber\\
     &H^{V}_{\frac{1}{2},0}=-\sqrt{\frac{2(M^2_{-}-q^2)}{3q^2}}\Big(F_{1}^{V}\frac{M_{+}M_{-}-q^2}{2M_2}-F_{2}^{V}\frac{(M^2_{+}-q^2)M_{-}}{2M_1M_2}+F_{3}^{V}\frac{|\vec{p'}|^2}{M_2}\Big),\nonumber\\
     &H^{V}_{\frac{1}{2},1}=\sqrt{\frac{M^2_{-}-q^2}{3}}\Big(F_{1}^{V}-F_{2}^{V}\frac{M^2_{+}-q^2}{M_1M_2}\Big),\nonumber\\
     &H^{V}_{\frac{3}{2},1}=-\sqrt{M^2_{-}-q^2}F_{1}^{V},	
     \end{eqnarray}
     and
     \begin{eqnarray}
     	&H^{A}_{\frac{1}{2},t}=\sqrt{\frac{2(M^2_{-}-q^2)}{3q^2}}(\frac{M^2_{+}-q^2}{2M_{1}M_{2}})\Big(F_{1}^{A}M_{1}+F_{2}^{A}M_{-}+F_{3}^{A}\frac{M_{+}M_{-}-q^2}{2M_1}+F_{4}^{A}\frac{q^{2}}{M_{1}}\Big),\nonumber\\
     &H^{A}_{\frac{1}{2},0}=\sqrt{\frac{2(M^2_{+}-q^2)}{3q^2}}\Big(F_{1}^{A}\frac{M_{+}M_{-}-q^2}{2M_2}+F_{2}^{A}\frac{(M^2_{-}-q^2)M_{+}}{2M_1M_2}+F_{3}^{A}\frac{|\vec{p'}|^2}{M_2}\Big),\nonumber\\
     &H^{A}_{\frac{1}{2},1}=\sqrt{\frac{M^2_{+}-q^2}{3}}\Big(F_{1}^{A}-F_{2}^{A}\frac{M^2_{-}-q^2}{M_1M_2}\Big),\nonumber\\
     &H^{A}_{\frac{3}{2},1}=\sqrt{M^2_{+}-q^2}F_{1}^{A},		
     \end{eqnarray}
   where $|\vec{P'}|=\frac{\lambda^{\frac{1}{2}}(M_{\Omega_{b}}^2,M_{\Omega^{*}_{c}}^2, q^2)}{2M_{\Omega_b}}$ and the other helicity amplitudes are given by $	H^{V,A}_{-\lambda_{2},-\lambda_{M}}=\mp H^{V,A}_{\lambda_{2},\lambda_{M}}$. Subsequently, the decay width of the two-body nonleptonic weak decay, $\Omega_{b}\rightarrow\Omega^{*}_{c}M$, for pseudoscalar (P) and vector (V) meson case, is obtained as \cite{Gutsche:2018utw},
  \begin{eqnarray}
  &&\Gamma(\frac{1}{2}\rightarrow\frac{3}{2})=\Gamma(\Omega_{b}\rightarrow\Omega^{*}_{c}M)=(\frac{G_F^2}{32\pi})\frac{|\vec{P'}|}{ M_{\Omega_{b}}^{2}}|V_{cb}V^*_{qq'}|^2C_{eff}^{2}f_{M}^2m_{Meson}^2\mathcal{H}_{N},\nonumber\\&&\mathcal{H}_{N}=\sum_{\lambda_2, \lambda_M}|H_{\lambda_2, \lambda_M}|^2,	
  \end{eqnarray} 
  in which $f_M$ and $m_{Meson}$ denote the decay constant and mass of the considered meson, respectively. The helicity components of the final baryon, $\Omega^{*}_{c}$, run over $\lambda_2=\pm\frac{1}{2}, \pm\frac{3}{2}$, the vector meson case has $\lambda_V=0, \pm1$; and the helicity of the pseudoscalar meson is $\lambda_P=t$, considering the constraint $|\lambda_2-\lambda_M|\le\frac{1}{2}$.
  
  The acquired form factors, employing the QCD‌ sum rules method, and helicity amplitudes enable us to determine the decay widths and branching ratios of the nonleptonic $\Omega_{b}\rightarrow\Omega^{*}_{c}(2766)M$ transitions as presented in Table \ref{decaywidth12}. 
   \begin{table}
  	\centering
  	\caption{Decay widths and branching ratios of the nonleptonic $\Omega_{b}\rightarrow\Omega^{*}_{c}(2766)M$ transitions with $M$ indicating a pseudoscalar or vector meson, considering uncertainties from both auxiliary and input QCD parameters.}
  	\begin{ruledtabular}
  		\begin{tabular}{ccc}
  			transition&$\Gamma$(GeV)&$Br$\\
  			\hline
  			$\Omega_{b}\rightarrow \Omega^{*}_{c}\pi^{-}$&$(1.00^{+0.22}_{-0.18})\times10^{-14}$&$(2.49^{+0.53}_{-0.43})\times10^{-2}$\\
  			$\Omega_{b}\rightarrow \Omega^{*}_{c}K^{-}$&$(8.59^{+1.85}_{-1.62})\times10^{-16}$&$(2.14^{+0.46}_{-0.40})\times10^{-3}$\\
  			$\Omega_{b}\rightarrow \Omega^{*}_{c}D^{-}$&$(4.64^{+0.99}_{-0.86})\times10^{-16}$&$(1.16^{+0.24}_{-0.23})\times10^{-3}$\\
  			$\Omega_{b}\rightarrow \Omega^{*}_{c}D_{s}^{-}$&$(1.11^{+0.23}_{-0.19})\times10^{-14}$&$(2.76^{+0.58}_{-0.47})\times10^{-2}$\\
  			$\Omega_{b}\rightarrow \Omega^{*}_{c}\rho^{-}$&$(2.89^{+0.59}_{-0.58})\times10^{-14}$&$(7.21^{+1.55}_{-1.34})\times10^{-2}$\\
  			$\Omega_{b}\rightarrow \Omega^{*}_{c}K^{*-}$&$(1.49^{+0.31}_{-0.34})\times10^{-15}$&$(3.72^{+0.76}_{-0.83})\times10^{-3}$\\
  			$\Omega_{b}\rightarrow \Omega^{*}_{c}D^{*-}$&$(1.04^{+0.19}_{-0.18})\times10^{-15}$&$(2.60^{+0.46}_{-0.40})\times10^{-3}$\\
  			$\Omega_{b}\rightarrow \Omega^{*}_{c}D_{s}^{*-}$&$(1.77^{+0.39}_{-0.37})\times10^{-14}$&$(4.41^{+0.97}_{-0.90})\times10^{-2}$\\
  		\end{tabular}
  	\end{ruledtabular}
  	\label{decaywidth12}
  \end{table} 
  A comparison of the responsible form factors for the semileptonic $\Omega_{b}\rightarrow \Omega^{*}_{c}\ell\bar{\nu}_{\ell}$ transition between this study (QCD sum rules) and light front approach has been provided in Table \ref{comparison ff semi omegab}. It is essential to note that light front approach follows different conventions for definitions of the form factors (see Ref. \cite{Chua:2019yqh}).  As is seen, although there are some consistencies between predictions of two studies on the values of some form factors,  in most cases they differ from each other. More theoretical studies should be carried out using different nonperturbative approaches.   We compare the obtained result of the decay width for the $\Omega_{b}\rightarrow \Omega^{*}_{c}\ell\bar{\nu}_{\ell}$ transition with the predictions of other theoretical approaches in Table \ref{comparison}. Our result is comparable with the findings of other methods with taking the uncertainties into account. A detailed investigation of semileptonic and nonleptonic weak decays of b-heavy baryons can improve our understanding of the hadronic structures, constrain the QCD parameters, and evaluate the SM predictions. In addition, considerable progress, in experiments, makes it more possible to explore the new allowed decay modes of b-heavy baryons in the near future.
  \begin{table}
	\centering
	\caption{comparison of the responsible form factors for the $\Omega_{b}\rightarrow \Omega_{c}^{*}\ell\bar{\nu}_{\ell}$ weak transition within QCD sum rules and light front approach.}
	\begin{ruledtabular}
		\begin{tabular}{|c|c|c|c|c|c|c|c|c|}
			&$F_{1}(q^{2})$&$F_{2}(q^{2})$&$F_{3}(q^{2})$&$F_{4}(q^{2})$&$G_{1}(q^{2})$&$G_{2}(q^{2})$&$G_{3}(q^{2})$&$G_{4}(q^{2})$\\
			\hline
			$\mathrm{this~ study}$&$0.043\pm0.001$&$-1.54\pm0.23$&$1.54\pm0.40$&$-0.77\pm0.28$&$-0.042\pm0.001$&$-0.59\pm0.25$&$-0.69\pm0.25$&$0.77\pm0.28$\\
			$\mathrm{light~front}$&-0.734&-0.300&1.06&0.033&0.526&0.088&-0.67&0.021\\
		\end{tabular}
	\end{ruledtabular}
	\label{comparison ff semi omegab}
\end{table}

   \begin{table}[h!]
  	\centering
  	\caption{Comparison of various theoretical results on the decay width of $\Omega_{b}\rightarrow \Omega^{*}_{c}(2766)\ell\bar{\nu}_{\ell}$ transition (in $10^{-14}$ GeV).}
  	\begin{ruledtabular}
  		\begin{tabular}{cccccccc}
  			transition&this work&\cite{Du:2011nj}&\cite{Ebert:2006rp}&\cite{Ivanov:1998ya}-I&\cite{Ivanov:1998ya}-II&\cite{Lu:2023rmq}&\cite{Zhang:2025pde}\\
  			\hline
  			$\Omega_{b}\rightarrow \Omega^{*}_{c}\ell\bar{\nu}_{\ell}$&$4.30^{+1.67}_{-1.72}$&$3.48$&$1.99$&$2.71$&$2.99$&$1.31$&$1.46\pm0.39$
  		\end{tabular}
  	\end{ruledtabular}
  	\label{comparison}
  \end{table} 
  \section{summary and conclusion}\label{conclusion}
  An extensive investigation of  b-heavy baryons provides us with a promising research area to explore the signals of new physics, CP violation, heavy quark effects on the hadronic structures, and QCD properties. Hence, the examination of bottom baryons, theoretically and experimentally, is of great interest. In the present work, we conducted a study of the semileptonic and nonleptonic weak decays of singly b-heavy baryons $\Omega^{*}_{b}$ and $\Omega_{b}$ with spin $\frac{3}{2}$ and $\frac{1}{2}$, respectively. We employed QCD sum rules method to obtain the responsible form factors of the  $\Omega^{*}_b\rightarrow\Omega_c\ell\bar{\nu}_{\ell}$ and $\Omega_b\rightarrow\Omega^*_c\ell\bar{\nu}_{\ell}$ semileptonic weak transitions. Utilizing the interpolating currents of the initial and final baryons as well as the OPE series up to dimension six, we derived the sum rules related to the transition form factors in terms of the hadronic, QCD, and auxiliary parameters. The specified auxiliary parameters provided us with the fit functions of the transition form factors in terms of $q^2$ in the whole physical region. The ultimate results were employed to determine the decay widths of the considered semileptonic weak transitions in all leptonic channels. We compared our finding for the decay width of the $\Omega_b\rightarrow\Omega^*_c\ell\bar{\nu}_{\ell}$ transition with the existing information in the literature. In addition, we presented the branching ratios related to the $\Omega_b$ baryon in different decay modes, taking the lifetime of this baryon from PDG. Subsequently, the obtained form factors utilized to compute the nonleptonic decay widths in different modes with emitting a pseudoscalar or vector meson.
  
  As previously remarked, a large number of singly b-heavy baryons have been observed in different experiments at various center of mass energies and integrated luminosities, for example, $\Omega_b$ baryon identified at $\sqrt{s}$=1.96 TeV and an integrated luminosity of 1.3 fb$^{-1}$. These days, concerning the great advances in experimental facilities such as LHC with $\sqrt{s}$=13.6 TeV and near future high-luminosity phase, it would be increasingly likely to observe and identify b-heavy baryons in various decay channels.
  
  Our findings indicate that in $\Omega_b$-decays sector, the nonleptonic $\Omega_{b}\rightarrow \Omega^{*}_{c}\pi^{-}$ and $\Omega_{b}\rightarrow \Omega^{*}_{c}D_{s}^{-}$ decays with emitting pseudoscalar mesons and 	$\Omega_{b}\rightarrow \Omega^{*}_{c}\rho^{-}$ and $\Omega_{b}\rightarrow \Omega^{*}_{c}D_{s}^{*-}$ decays with vector mesons would be accessible candidates for exploring in various experiments such as LHCb or Belle. In addition, regarding the branching ratios of the semileptonic $\Omega_{b}\rightarrow \Omega_{c}^{*}\ell \bar{\nu}_{\ell}$ decays, it is obvious that the considered decays in the $e$ and $\mu$ lepton channels are more probable to be identified. Moreover, in the $\Omega^{*}_b$-decays sector, although its dominant decay channel is strong, in accordance with our previous study\cite{Khajouei:2024frw}, the semileptonic $b\rightarrow c$ weak transitions are expected to have acceptable contributions to the total width, therefore, the semileptonic $\Omega^{*}_{b}\rightarrow \Omega_{c}e\bar{\nu}_{e}$ and $\Omega^{*}_{b}\rightarrow \Omega_{c}\mu\bar{\nu}_{\mu}$ and nonleptonics $\Omega^{*}_{b}\rightarrow \Omega_{c}\pi^{-}$, $\Omega^{*}_{b}\rightarrow \Omega_{c}D_{s}^{-}$, $\Omega^{*}_{b}\rightarrow \Omega_{c}\rho^{-}$ and $\Omega^{*}_{b}\rightarrow \Omega_{c}D_{s}^{*-}$ are more expected to be observed and identified at different experiments.
  Considering the significant advances in experiments, it is more likely to identify the more expected allowed decay channels of heavy baryons. We hope the findings presented in this study provide valuable information for future experiments aimed at exploring new decay channels and identifying the heavy baryons in order to have a remarkable insight into the internal structure of b-heavy baryons and search for new physics effects.
   
   \appendix
   \section{THE CORRELATION FUNCTION OF‌ THE SEMILEPTONIC $\Omega_b\rightarrow\Omega^*_c\ell\bar{\nu}_{\ell}$ TRANSITION}\label{AppA}
   In this appendix, we present the physical and QCD  representations of the correlation function associated with the semileptonic $\Omega_b\rightarrow\Omega^*_c\ell\bar{\nu}_{\ell}$ transition.  First, we present the corresponding transition matrix elements in physical side:  
	  \begin{eqnarray}\label{ff vector omegab}
	 	&&\langle\Omega_{c}^*(p^{\prime},s^{\prime})|J_{\mu}^{V}|\Omega_{b}(p,s)\rangle=
	 	\langle\Omega_{c}^*(p^{\prime},s^{\prime})|\bar{c}\gamma_{\mu}b|\Omega_{b}(p,s)\rangle\nonumber\\
	 	&&=\bar{u}_{\Omega_{c}^*}^{\alpha}(p^{\prime},s^{\prime})\Big[g_{\mu\alpha}F_{1}(q^{2})+\gamma_{\mu}\frac{p_{\alpha}}{M_{\Omega_{b}}}F_{2}(q^{2})+\frac{p_{\alpha}p'_{\mu}}{M_{\Omega_{b}}^{2}}F_{3}(q^{2})+\frac{p_{\alpha}q_{\mu}}{M_{\Omega_{b}}^{2}}F_{4}(q^{2})\Big]\gamma_{5}u_{\Omega_{b}}(p,s),\nonumber
	 \end{eqnarray}      
		 \begin{eqnarray}\label{ff axialvector omegab}
		&&\langle\Omega_{c}^*(p^{\prime},s^{\prime})|J_{\mu}^{A}|\Omega_{b}(p,s)\rangle=
		\langle\Omega_{c}^*(p^{\prime},s^{\prime})|\bar{c}\gamma_{\mu}\gamma_{5}b|\Omega_{b}(p,s)\rangle\nonumber\\
		&&=\bar{u}_{\Omega_{c}^*}^{\alpha}(p^{\prime},s^{\prime})\,\Big[g_{\mu\alpha}G_{1}(q^{2})+\gamma_{\mu}\frac{p_{\alpha}}{M_{\Omega_{b}}}\,G_{2}(q^{2})+\frac{p_{\alpha}p'_{\mu}}{M_{\Omega_{b}}^{2}}G_{3}(q^{2})+\frac{p_{\alpha}q_{\mu}}{M_{\Omega_{b}}^{2}}G_{4}(q^{2})\Big]u_{\Omega_{b}}(p,s).
	\end{eqnarray}
Using the standard procedures explained in the text,  for the physical side of the $\Omega_b\rightarrow\Omega^{*}_c\ell\bar{\nu}_{\ell}$ transition, we get, 
   \begin{eqnarray}\label{physicalside2}
   	&&\mathbf{\widehat{B}}\Pi_{\mu\nu}^{Phys.}(p,p^{\prime},q^2)=\lambda_{\Omega_{b}}\lambda_{\Omega^{*}_{c}}e^{-\frac{m_{\Omega_{b}}^{2}}{M^{2}}}e^{-\frac{m_{\Omega^{*}_{c}}^{2}}{M\prime^{2}}}\Bigg[F_{1}\Big(-m_{\Omega_{b}}m_{\Omega^{*}_{c}}g_{\mu\nu}\gamma_{5}+m_{\Omega^{*}_{c}}g_{\mu\nu}\slashed{p}\gamma_{5}-m_{\Omega_{b}}g_{\mu\nu}\slashed{p}^{\prime}\gamma_{5}+g_{\mu\nu}\slashed{p}\slashed{p}^{\prime}\gamma_{5}\Big)+\notag\\&&F_{2}\Big(\frac{2}{m_{\Omega_{b}}}p_{\mu}p_{\nu}\slashed{p}'\gamma_{5}-p_{\mu}\slashed{p}'\gamma_{\nu}\gamma_{5}-\frac{m_{\Omega_{c}^{*}}}{m_{\Omega_{b}}}p_{\mu}\slashed{p}\gamma_{\nu}\gamma_{5}+\frac{1}{m_{\Omega_{b}}}p_{\mu}\slashed{p}\slashed{p}^{\prime}\gamma_{\nu}\gamma_{5}\Big)+F_{3}\Big(-\frac{m_{\Omega_{c}^{*}}}{m_{\Omega_{b}}}p_{\mu}p^{\prime}_{\nu}\gamma_{5}-\frac{1}{m_{\Omega_{b}}}p_{\mu}p^{\prime}_{\nu}\slashed{p}'\gamma_{5}+\notag\\&&\frac{m_{\Omega_{c}^{*}}}{m_{\Omega_{b}}^{2}}p_{\mu}p^{\prime}_{\nu}\slashed{p}\gamma_{5}-\frac{1}{m_{\Omega_{b}}^{2}}p_{\mu}p^{\prime}_{\nu}\slashed{p}\slashed{p}^{\prime}\gamma_{5}\Big)+F_{4}\Big(\frac{1}{m_{\Omega_{b}}}p_{\mu}p^{\prime}_{\nu}\slashed{p}^{\prime}\gamma_{5}-\frac{1}{m_{\Omega_{b}}}p_{\mu}p_{\nu}\slashed{p}^{\prime}\gamma_{5}+\frac{1}{m_{\Omega_{b}}^{2}}p_{\mu}p^{\prime}_{\nu}\slashed{p}\slashed{p}^{\prime}\gamma_{5}-\frac{1}{m_{\Omega_{b}}^{2}}p_{\mu}p_{\nu}\slashed{p}\slashed{p}^{\prime}\gamma_{5}\Big)\notag\\&&-G_{1}\Big(-m_{\Omega_{b}}m_{\Omega^{*}_{c}}g_{\mu\nu}-m_{\Omega^{*}_{c}}g_{\mu\nu}\slashed{p}-m_{\Omega_{b}}g_{\mu\nu}\slashed{p}^{\prime}+g_{\mu\nu}\slashed{p}\slashed{p}^{\prime}\Big)-G_{2}\Big(-\frac{2}{m_{\Omega_{b}}}p_{\mu}p_{\nu}\slashed{p}'-p_{\mu}\gamma_{\nu}\slashed{p}'+\frac{m_{\Omega_{c}^{*}}}{m_{\Omega_{b}}}p_{\mu}\gamma_{\nu}\slashed{p}\notag\\&&-\frac{1}{m_{\Omega_{b}}}p_{\mu}\slashed{p}\slashed{p}^{\prime}\gamma_{\nu}\Big)-G_{3}\Big(-\frac{m_{\Omega_{c}^{*}}}{m_{\Omega_{b}}}p_{\mu}p^{\prime}_{\nu}-\frac{1}{m_{\Omega_{b}}}p_{\mu}p^{\prime}_{\nu}\slashed{p}'-\frac{m_{\Omega_{c}^{*}}}{m_{\Omega_{b}}^{2}}p_{\mu}p^{\prime}_{\nu}\slashed{p}+\frac{1}{m_{\Omega_{b}}^{2}}p_{\mu}p^{\prime}_{\nu}\slashed{p}\slashed{p}^{\prime}\Big)-G_{4}\Big(-\frac{1}{m_{\Omega_{b}}}p_{\mu}p_{\nu}\slashed{p}^{\prime}\notag\\&&+\frac{1}{m_{\Omega_{b}}}p_{\mu}p^{\prime}_{\nu}\slashed{p}^{\prime}-\frac{1}{m_{\Omega_{b}}^{2}}p_{\mu}p^{\prime}_{\nu}\slashed{p}\slashed{p}^{\prime}+\frac{1}{m_{\Omega_{b}}^{2}}p_{\mu}p_{\nu}\slashed{p}\slashed{p}^{\prime}\Big)\Bigg]+\cdots.
   \end{eqnarray}
    In addition, by applying Wick's theorem, contracting the related quark fields, and in terms of the heavy and light quark propagators, the QCD side of the correlation function for the semileptonic $\Omega_b\rightarrow\Omega^*_c\ell\bar{\nu}_{\ell}$ transition is derived as,
    \begin{eqnarray}\label{qcdside2}
    	&&\Pi_{\mu\nu}^{QCD}(p,p^{\prime },q^{2})=i^{2}\int\mathrm{d}^{4}x e^{-ipx}\int\mathrm{d}^{4}y e^{ip^{\prime}y}\frac{1}{\sqrt{2}}\frac{1}{\sqrt{3}}\epsilon^{abc}\epsilon^{a^{\prime} b^{\prime} c^{\prime}}\Bigg\{-S^{ci}_{c}(y)\gamma_{\nu}(1-\gamma_5)S^{ib'}_{b}(-x)\tilde{S}^{aa'}_{s}(y-x)\gamma_{\mu}\nonumber\\&&S^{bc'}_{s}(y-x)\gamma_5+S^{ci}_{c}(y)\gamma_{\nu}(1-\gamma_5)S^{ib'}_{b}(-x)\tilde{S}^{ba'}_{s}(y-x)\gamma_{\mu}S^{ac'}_{s}(y-x)\gamma_5-\beta S^{ci}_{c}(y)\gamma_{\nu}(1-\gamma_5)S^{ib'}_{b}(-x)\gamma_5\nonumber\\&&\tilde{S}^{aa'}_{s}(y-x)\gamma_{\mu}S^{bc'}_{s}(y-x)+\beta S^{ci}_{c}(y)\gamma_{\nu}(1-\gamma_5)S^{ib'}_{b}(-x)\gamma_5\tilde{S}^{ba'}_{s}(y-x)\gamma_{\mu}S^{ac'}_{s}(y-x)+S^{ci}_{c}(y)\gamma_{\nu}(1-\gamma_5)\nonumber\\&&S^{ia'}_{b}(-x)\tilde{S}^{ab'}_{s}(y-x)\gamma_{\mu}S^{bc'}_{s}(y-x)\gamma_5-S^{ci}_{c}(y)\gamma_{\nu}(1-\gamma_5)S^{ia'}_{b}(-x)\tilde{S}^{bb'}_{s}(y-x)\gamma_{\mu}S^{ac'}_{s}(y-x)\gamma_5+\beta S^{ci}_{c}(y)\nonumber\\&&\gamma_{\nu}(1-\gamma_5)S^{ia'}_{b}(-x)\gamma_5\tilde{S}^{ab'}_{s}(y-x)\gamma_{\mu}S^{bc'}_{s}(y-x)-\beta S^{ci}_{c}(y)\gamma_{\nu}(1-\gamma_5)S^{ia'}_{b}(-x)\gamma_5\tilde{S}^{bb'}_{s}(y-x)\gamma_{\mu}S^{ac'}_{s}(y-x)+\nonumber\\&&S^{cc'}_{s}(y-x)\gamma_5Tr[\gamma_{\mu}S^{bi}_{c}(y)\gamma_{\nu}(1-\gamma_5)S^{ib'}_{b}(-x)\tilde{S}^{aa'}_{s}(y-x)]+S^{ca'}_{s}(y-x)\tilde{S}^{ib'}_{b}(-x)(1-\gamma_5)\gamma_{\nu}\tilde{S}^{bi}_{c}(y)\gamma_{\mu}\nonumber\\&&S^{ac'}_{s}(y-x)\gamma_5+\beta S^{cc'}_{s}(y-x)Tr[\gamma_{\mu}S^{bi}_{c}(y)\gamma_{\nu}(1-\gamma_5)S^{ib'}_{b}(-x)\gamma_5\tilde{S}^{aa'}_{s}(y-x)]+\beta S^{ca'}_{s}(y-x)\gamma_5\tilde{S}^{ib'}_{b}(-x)\nonumber\\&&(1-\gamma_5)\gamma_{\nu}\tilde{S}^{bi}_{c}(y)\gamma_{\mu}S^{ac'}_{s}(y-x)-S^{cc'}_{s}(y-x)\gamma_5Tr[\gamma_{\mu}S^{bi}_{c}(y)\gamma_{\nu}(1-\gamma_5)S^{ia'}_{b}(-x)\tilde{S}^{ab'}_{s}(y-x)]-S^{cb'}_{s}(y-x)\nonumber\\&&\tilde{S}^{ia'}_{b}(-x)(1-\gamma_5)\gamma_{\nu}\tilde{S}^{bi}_{c}(y)\gamma_{\mu}S^{ac'}_{s}(y-x)\gamma_5-\beta S^{cc'}_{s}(y-x)Tr[\gamma_{\mu}S^{bi}_{c}(y)\gamma_{\nu}(1-\gamma_5)S^{ia'}_{b}(-x)\gamma_5\tilde{S}^{ab'}_{s}(y-x)]\nonumber\\&&-\beta S^{cb'}_{s}(y-x)\gamma_5\tilde{S}^{ia'}_{b}(-x)(1-\gamma_5)\gamma_{\nu}\tilde{S}^{bi}_{c}(y)\gamma_{\mu}S^{ac'}_{s}(y-x)+S^{cc'}_{s}(y-x)\gamma_5Tr[\gamma_{\mu}S^{ba'}_{s}(y-x)\tilde{S}^{ib'}_{b}(-x)(1-\gamma_5)\nonumber\\&&\gamma_{\nu}\tilde{S}^{ai}_{c}(y)]-S^{ca'}_{s}(y-x)\tilde{S}^{ib'}_{b}(-x)(1-\gamma_5)\gamma_{\nu}\tilde{S}^{ai}_{c}(y)\gamma_{\mu}S^{bc'}_{s}(y-x)\gamma_5+\beta S^{cc'}_{s}(y-x)Tr[\gamma_{\mu}S^{ba'}_{s}(y-x)\gamma_5\tilde{S}^{ib'}_{b}(-x)\nonumber\\&&(1-\gamma_5)\gamma_{\nu}\tilde{S}^{ai}_{c}(y)]-\beta S^{ca'}_{s}(y-x)\gamma_5\tilde{S}^{ib'}_{b}(-x)(1-\gamma_5)\gamma_{\nu}\tilde{S}^{ai}_{c}(y)\gamma_{\mu}S^{bc'}_{s}(y-x)-S^{cc'}_{s}(y-x)\gamma_5\nonumber\\&&Tr[\gamma_{\mu}S^{bb'}_{s}(y-x)\tilde{S}^{ia'}_{b}(-x)(1-\gamma_5)\gamma_{\nu}\tilde{S}^{ai}_{c}(y)]+S^{cb'}_{s}(y-x)\tilde{S}^{ia'}_{b}(-x)(1-\gamma_5)\gamma_{\nu}\tilde{S}^{ai}_{c}(y)\gamma_{\mu}S^{bc'}_{s}(y-x)\gamma_5-\beta\nonumber\\&& S^{cc'}_{s}(y-x)Tr[\gamma_{\mu}S^{bb'}_{s}(y-x)\gamma_5\tilde{S}^{ia'}_{b}(-x)(1-\gamma_5)\gamma_{\nu}\tilde{S}^{ai}_{c}(y)]+\beta S^{cb'}_{s}(y-x)\gamma_5\tilde{S}^{ia'}_{b}(-x)(1-\gamma_5)\gamma_{\nu}\tilde{S}^{ai}_{c}(y)\nonumber\\&&\gamma_{\mu}S^{bc'}_{s}(y-x)\Bigg\}.
    \end{eqnarray}
    \section{THE EXPRESSIONS OF‌ THE‌ SUM RULES FOR THE FORM FACTORS}\label{AppB}
    The sum rules expressions for the form factors $F_{1}, F_{2}, F_{3}, F_{4}, G_{1}, G_{2}, G_{3}$, and $G_{4}$, related to the $\Omega^{*}_b\rightarrow\Omega_c\ell\bar{\nu}_{\ell}$ semileptonic transition, are presented in this appendix,
     \begin{eqnarray}
    	F_{1}(q^{2})&=&-\frac{e^{m_{\Omega_{b}^{*}}^{2}/M_{1}^{2}}e^{m_{\Omega_{c}}^{2}/M_{2}^{2}}}{m_{\Omega_{b}^{*}}\lambda_{\Omega_{b}^{*}}\lambda_{\Omega_{c}}}\Bigg\{\Bigg[\int_{(m_{b}+2m_{s})^{2}}^{s_{0}}\mathrm{d}s\int_{(m_{c}+2m_{s})^{2}}^{s_{0}^{\prime}}\mathrm{d}s^{\prime}e^{-s/M_{1}^{2}}e^{-s^{\prime}/M_{2}^{2}}\bigg(\rho_{g_{\mu\nu}\slashed{p}'\gamma_{5}}^{Pert.}(s,s^{\prime},q^{2})\notag\\&+&\rho_{g_{\mu\nu}\slashed{ p}'\gamma_{5}}^{Dim-3}(s,s^{\prime},q^{2})\bigg)\Bigg]+\mathbf{\widehat{B}}\big[\Gamma_{g_{\mu\nu}\slashed{p}'\gamma_{5}}^{Dim-6}(p^{2},p^{\prime2},q^{2})\big]\Bigg\}
    \end{eqnarray}
     \begin{eqnarray}
    	F_{2}(q^{2})&=&\frac{m_{\Omega_{c}}e^{m_{\Omega_{b}^{*}}^{2}/M_{1}^{2}}e^{m_{\Omega_{c}}^{2}/M_{2}^{2}}}{\lambda_{\Omega_{b}^{*}}\lambda_{\Omega_{c}}}\Bigg\{\Bigg[\int_{(m_{b}+2m_{s})^{2}}^{s_{0}}\mathrm{d}s\int_{(m_{c}+2m_{s})^{2}}^{s_{0}^{\prime}}\mathrm{d}s^{\prime}e^{-s/M_{1}^{2}}e^{-s^{\prime}/M_{2}^{2}}\bigg(\rho_{p^{\prime}_{\nu}\gamma_{\mu}\slashed{p}\slashed{p}^{\prime}\gamma_{5}}^{Pert.}(s,s^{\prime},q^{2})\bigg)\Bigg]\notag\\&+&\mathbf{\widehat{B}}\big[\Gamma_{p^{\prime}_{\nu}\gamma_{\mu}\slashed{p}\slashed{p}^{\prime}\gamma_{5}}^{Dim-6}(p^{2},p^{\prime2},q^{2})\big]\Bigg\}
    \end{eqnarray}
     \begin{eqnarray}
    	F_{3}(q^{2})&=&-\frac{m_{\Omega_{c}}^2e^{m_{\Omega_{b}^{*}}^{2}/M_{1}^{2}}e^{m_{\Omega_{c}}^{2}/M_{2}^{2}}}{m_{\Omega_{b}^{*}}\lambda_{\Omega_{b}^{*}}\lambda_{\Omega_{c}}}\Bigg\{\Bigg[\int_{(m_{b}+2m_{s})^{2}}^{s_{0}}\mathrm{d}s\int_{(m_{c}+2m_{s})^{2}}^{s_{0}^{\prime}}\mathrm{d}s^{\prime}e^{-s/M_{1}^{2}}e^{-s^{\prime}/M_{2}^{2}}\bigg(\rho_{p_{\mu}p^{\prime}_{\nu}\slashed{p}^{\prime}\gamma_{5}}^{Pert.}(s,s^{\prime},q^{2})\bigg)\Bigg]\notag\\&+&\mathbf{\widehat{B}}\big[\Gamma_{p_{\mu}p^{\prime}_{\nu}\slashed{p}^{\prime}\gamma_{5}}^{Dim-6}(p^{2},p^{\prime2},q^{2})\big]\Bigg\}-\frac{m_{\Omega_{c}}^2e^{m_{\Omega_{b}^{*}}^{2}/M_{1}^{2}}e^{m_{\Omega_{c}}^{2}/M_{2}^{2}}}{\lambda_{\Omega_{b}^{*}}\lambda_{\Omega_{c}}}\Bigg\{\int_{(m_{b}+2m_{s})^{2}}^{s_{0}}\mathrm{d}s\int_{(m_{c}+2m_{s})^{2}}^{s_{0}^{\prime}}\mathrm{d}s^{\prime}e^{-s/M_{1}^{2}}e^{-s^{\prime}/M_{2}^{2}}\notag\\&\bigg(&\rho_{p^{\prime}_{\mu}p^{\prime}_{\nu}\slashed{p}\slashed{p}^{\prime}\gamma_{5}}^{Pert.}(s,s^{\prime},q^{2})\bigg)\Bigg\}
    \end{eqnarray}	
     \begin{eqnarray}
    	F_{4}(q^{2})&=&\frac{m_{\Omega_{c}}^2e^{m_{\Omega_{b}^{*}}^{2}/M_{1}^{2}}e^{m_{\Omega_{c}}^{2}/M_{2}^{2}}}{\lambda_{\Omega_{b}^{*}}\lambda_{\Omega_{c}}}\Bigg\{\int_{(m_{b}+2m_{s})^{2}}^{s_{0}}\mathrm{d}s\int_{(m_{c}+2m_{s})^{2}}^{s_{0}^{\prime}}\mathrm{d}s^{\prime}e^{-s/M_{1}^{2}}e^{-s^{\prime}/M_{2}^{2}}\bigg(\rho_{p^{\prime}_{\mu}p^{\prime}_{\nu}\slashed{p}\slashed{p}^{\prime}\gamma_{5}}^{Pert.}(s,s^{\prime},q^{2})\bigg)\Bigg\}
    \end{eqnarray}
     \begin{eqnarray}
    	G_{1}(q^{2})&=&\frac{e^{m_{\Omega_{b}^{*}}^{2}/M_{1}^{2}}e^{m_{\Omega_{c}}^{2}/M_{2}^{2}}}{m_{\Omega_{c}}\lambda_{\Omega_{b}^{*}}\lambda_{\Omega_{c}}}\Bigg\{\Bigg[\int_{(m_{b}+2m_{s})^{2}}^{s_{0}}\mathrm{d}s\int_{(m_{c}+2m_{s})^{2}}^{s_{0}^{\prime}}\mathrm{d}s^{\prime}e^{-s/M_{1}^{2}}e^{-s^{\prime}/M_{2}^{2}}\bigg(\rho_{g_{\mu\nu}\slashed{p}}^{Pert.}(s,s^{\prime},q^{2})\notag\\&+&\rho_{g_{\mu\nu}\slashed{ p}}^{Dim-3}(s,s^{\prime},q^{2})\bigg)\Bigg]+\mathbf{\widehat{B}}\big[\Gamma_{g_{\mu\nu}\slashed{p}}^{Dim-6}(p^{2},p^{\prime2},q^{2})\big]\Bigg\}
    \end{eqnarray}
    \begin{eqnarray}
    	G_{2}(q^{2})&=&-\frac{m_{\Omega_{c}}e^{m_{\Omega_{b}^{*}}^{2}/M_{1}^{2}}e^{m_{\Omega_{c}}^{2}/M_{2}^{2}}}{m_{\Omega_{b}^{*}}\lambda_{\Omega_{b}^{*}}\lambda_{\Omega_{c}}}\Bigg\{\Bigg[\int_{(m_{b}+2m_{s})^{2}}^{s_{0}}\mathrm{d}s\int_{(m_{c}+2m_{s})^{2}}^{s_{0}^{\prime}}\mathrm{d}s^{\prime}e^{-s/M_{1}^{2}}e^{-s^{\prime}/M_{2}^{2}}\bigg(\rho_{p^{\prime}_{\nu}\gamma_{\mu}\slashed{p}^{\prime}}^{Pert.}(s,s^{\prime},q^{2})\bigg)\Bigg]\notag\\&+&\mathbf{\widehat{B}}\big[\Gamma_{p^{\prime}_{\nu}\gamma_{\mu}\slashed{p}^{\prime}}^{Dim-6}(p^{2},p^{\prime2},q^{2})\big]\Bigg\}
    \end{eqnarray}
    \begin{eqnarray}
    	G_{3}(q^{2})&=&\frac{m_{\Omega_{c}}e^{m_{\Omega_{b}^{*}}^{2}/M_{1}^{2}}e^{m_{\Omega_{c}}^{2}/M_{2}^{2}}}{\lambda_{\Omega_{b}^{*}}\lambda_{\Omega_{c}}}\Bigg\{\Bigg[\int_{(m_{b}+2m_{s})^{2}}^{s_{0}}\mathrm{d}s\int_{(m_{c}+2m_{s})^{2}}^{s_{0}^{\prime}}\mathrm{d}s^{\prime}e^{-s/M_{1}^{2}}e^{-s^{\prime}/M_{2}^{2}}\bigg(\rho_{p_{\mu}p^{\prime}_{\nu}\slashed{p}}^{Pert.}(s,s^{\prime},q^{2})\bigg)\Bigg]\notag\\&+&\mathbf{\widehat{B}}\big[\Gamma_{p_{\mu}p^{\prime}_{\nu}\slashed{p}}^{Dim-6}(p^{2},p^{\prime2},q^{2})\big]\Bigg\}-\frac{m_{\Omega_{c}}^2e^{m_{\Omega_{b}^{*}}^{2}/M_{1}^{2}}e^{m_{\Omega_{c}}^{2}/M_{2}^{2}}}{\lambda_{\Omega_{b}^{*}}\lambda_{\Omega_{c}}}\Bigg\{\int_{(m_{b}+2m_{s})^{2}}^{s_{0}}\mathrm{d}s\int_{(m_{c}+2m_{s})^{2}}^{s_{0}^{\prime}}\mathrm{d}s^{\prime}e^{-s/M_{1}^{2}}e^{-s^{\prime}/M_{2}^{2}}\notag\\&\bigg(&\rho_{p^{\prime}_{\mu}p^{\prime}_{\nu}\slashed{p}\slashed{p}^{\prime}}^{Pert.}(s,s^{\prime},q^{2})\bigg)\Bigg\}
    \end{eqnarray}
    \begin{eqnarray}
    	G_{4}(q^{2})&=&\frac{m_{\Omega_{c}}^2e^{m_{\Omega_{b}^{*}}^{2}/M_{1}^{2}}e^{m_{\Omega_{c}}^{2}/M_{2}^{2}}}{\lambda_{\Omega_{b}^{*}}\lambda_{\Omega_{c}}}\Bigg\{\int_{(m_{b}+2m_{s})^{2}}^{s_{0}}\mathrm{d}s\int_{(m_{c}+2m_{s})^{2}}^{s_{0}^{\prime}}\mathrm{d}s^{\prime}e^{-s/M_{1}^{2}}e^{-s^{\prime}/M_{2}^{2}}\bigg(\rho_{p^{\prime}_{\mu}p^{\prime}_{\nu}\slashed{p}\slashed{p}^{\prime}}^{Pert.}(s,s^{\prime},q^{2})\bigg)\Bigg\}
    \end{eqnarray}
    where the explicit representations of $\rho(s,s^{\prime},q^{2})$ and $\Gamma(p^{2},p^{\prime2},q^{2})$ are given in the following appendix.
    \section{ PERTURBATIVE AND NONPERTURBATIVE CONTRIBUTIONS}\label{AppC}
    The explicit expressions of the spectral densities, $\rho_{i}(s,s^{\prime},q^{2})$, and functions $\Gamma_{i}(p^{2},p^{\prime2},q^{2})$ for the $\Omega^{*}_b\rightarrow\Omega_c\ell\bar{\nu}_{\ell}$ semileptonic weak transition, as an example, for the structure $g_{\mu\nu}\slashed {p}'\gamma_5$ are presented as,
    \begin{eqnarray}
    &&\rho_{g_{\mu\nu}\slashed {p}'\gamma_5}^{Pert.}(s,s^{\prime},q^{2})=\int_{0}^{1}\mathrm{d}u\int_{0}^{1-u}\mathrm{d}v\frac{1}{512\sqrt{3}\pi^{4}F^{2}}\nonumber\\&&\Bigg\{\Bigg(L(-16m_{b}u- 
    3m_sF(-16+5\beta+16(7+3\beta)v)+5\beta(m_{b}u+m_{c}v)+ 
    16v(3m_{c}+6m_{b}F-4m_{c}(u+v)))-\nonumber\\&&F\Big(96(-1+\beta)m_{c}m_{s}^{2}v + m_{c}v\big(s(32+13\beta-32u)u+(-q^{2}+s+s')(16+5\beta-32u)v- 
    32s'v^{2}\big)+\nonumber\\&&m_{b}\big(-96(-1+\beta)m_{s}^{2}(-1+v)+48(3+\beta) m_{c}m_{s}v+ 
    \beta u(13su+5(-q^{2}+s+s')v)+ 
    16v(u(q^{2}-s-s'+2su)-\nonumber\\&&2s'v+2(-q^{2}+s+s')uv+ 
    2s'v^{2})\big)-3m_{s}(v(-((-16+5\beta)q^{2}(-1+u))- 
    q^{2}(-16+5\beta+16(3+\beta)u)v+\nonumber\\&& 
    s'F(-16+5\beta+16(3+\beta)v))+ 
    s F((-16+5\beta)v+ 
    u(16+29\beta+16(3+\beta)v)))\Big)\Bigg)\Bigg\}\Theta[L(s,s^{\prime},q^{2})],
    \end{eqnarray}
    \begin{eqnarray}
    \rho_{g_{\mu\nu}\slashed {p}'\gamma_5}^{Dim-3}(s,s^{\prime},q^{2})=\int_{0}^{1}\mathrm{d}u\int_{0}^{1-u}\mathrm{d}v\frac{-\langle \bar{s}s\rangle}{64\sqrt{3}\pi^{2}}\Bigg(\Big(-16+5\beta+16(7+3\beta)v\Big)\Bigg)\Theta[L(s,s^{\prime},q^{2})],
    \end{eqnarray}
   	\begin{equation}
   	\rho_{g_{\mu\nu}\slashed{p}'\gamma_5}^{Dim-4}(s,s^{\prime},q^{2})=0,
    \end{equation}
   	\begin{equation}
   	\rho_{g_{\mu\nu}\slashed{p}'\gamma_5}^{Dim-5}(s,s^{\prime},q^{2})=0,
    \end{equation}
    \begin{eqnarray}
    &&\mathbf{\widehat{B}}\big[\Gamma_{g_{\mu\nu}\slashed{p}'\gamma_{5}}^{Dim-6}(p^{2},p^{\prime2},q^{2})\big]=-\frac{1}{192\sqrt{3}M_{1}^{4}M_{2}^{4}}e^{-\frac{m_b^2}{M_1^{2}}}e^{-\frac{m_c^2}{M_2^{2}}}\Bigg(13\beta M_2^{4}m_b^{3}m_{s}^{2}+ 
    M_1^{2}M_2^{2}m_{b}m_s(-2M_2^{2}((16+29\beta)m_b+\nonumber\\&&13\beta m_s)+ 
    m_s((-16+5\beta)m_b^{2}+(32+13\beta)m_b m_c+(-16+ 
    5\beta)(m_c^{2}-q^{2})))+ 
    M_1^{4}\Big(M_2^{4}(64(-1+\beta)m_b+32m_s+\nonumber\\&&58\beta m_{s})+ 
    m_{c}m_s^{2}((16+5\beta)m_b^{2}- 
    32m_{b} m_{c}+(16+5\beta)(m_c^{2}-q^2))- 
    M_2^{2}m_s(2(-16+5\beta)m_b^{2}+ 
    32(3+\beta)m_b m_c+\nonumber\\&&(16+5\beta)m_b m_{s}+ 
    2m_c(-16m_c+5\beta m_{c}+8m_{s}+9\beta m_s)+ 
    2(16-5\beta)q^2)\Big)\Bigg)\langle \bar{s}s\rangle^{2}
    \end{eqnarray}	
    	where,
    \begin{eqnarray}
    	&&L(s,s^{\prime},q^{2})=-m_{b}^{2}u+su-su^{2}-m_{c}^{2}v+s'v-s u v-s'u v+q^{2}u v-s'v^{2},\nonumber\\&&F=-1+u+v,
    \end{eqnarray}
    and $\Theta[...]$ represents the unit step function.
   
      
  \end{document}